\documentclass{SCGEcontent}
\usepackage{titletoc}
\begin{document}

\begin{picture}(0,0){\rm
\put(0,-20){\makebox[160truemm][l]{\bf {\sanhao\raisebox{2pt}{.}}
Invited Review  {\sanhao\raisebox{1.5pt}{.}}}}}
\put(0,-34){\jiuwuhao {\textcolor[rgb]{0.5,0.5,0.5}{\sf Special Topic: the Next Detectors for Gravitational Wave Astronomy
}}}
\end{picture}

\def\bm{\boldsymbol}

\def\dl{\displaystyle}
\def\du{\end{document}}
\def\d{{\rm d}}
\def\e{{\rm e}}
\def\r{{\bm r}}
\def\P{{\bm P}}
\def\A{{\bm A}}
\def\k{{\bm k}}
\def\Q{{\bm Q}}

\def\cp#1{\mathbf{#1}}

\Year{2015} %
\Month{December} %
\Vol{X} 
\No{X} 
\BeginPage{1} 
\EndPage{11} 
\AuthorMark{{\rm Mitrofanov V P}, et al.}  
\AuthorMarkCite{{\rm Mitrofanov V P, Chao S, Pan H W}, et al.} 
\DOI{10.1007/s11433-015-5738-8} 
\ArtNo{X}

\title{Technology for the next gravitational wave detectors\!
\footnotemark[2]\footnotetext[2]{Sect. 1 is contributed by MITROFANOV Valery P. (email: mitr@hbar.phys.msu.ru); sect. 2 is contributed by CHAO Shiuh, PAN Huang-Wei, KUO Ling-Chi (corresponding author, CHAO Shiuh,
email: schao@ee.nthu.edu.tw); sect. 3 is contributed by COLE Garrett (email: garrett.cole@crystallinemirrors.com); sect. 4 is contributed by DEGALLAIX Jerome (email: j.degallaix@lma.in2p3.fr); sect. 5 is contributed by WILLKE Benno (email: benno.willke@aei.mpg.de)}
}

\author[1]{MITROFANOV Valery P.}{}
\author[2]{CHAO Shiuh}{}
\author[2]{PAN Huang-Wei}{}
\author[2]{KUO Ling-Chi}{}
\author[\!3,4,5]{\vspace*{1.3mm}\\COLE Garrett }{}
\author[6]{DEGALLAIX Jerome}{}
\author[7]{WILLKE Benno}{}

\address[{\rm1}]{Faculty of Physics, Moscow State University, Moscow 119991, Russia;}
\address[{\rm2}]{Institute of Photonics Technologies, ``National'' Tsing Hua University, Hsinchu 30013, China;}
\address[{\rm3}]{Crystalline Mirror Solutions LLC, 114 E Haley St., Suite G, Santa Barbara, CA 93101, USA;}
\address[{\rm4}]{Crystalline Mirror Solutions GmbH, Seestadtstr. 27, Top 1.05, A-1220 Vienna, Austria;}
\address[{\rm5}]{Vienna Center for Quantum Science and Technology (VCQ), Faculty of Physics, University of Vienna, Boltzmanngasse 5, Vienna A-1090, Austria;}
\address[{\rm6}]{Laboratoire des Mat\'eriaux Avanc\'es, 7 Avenue Pierre de Coubertin, Villeurbanne 69100, France;}
\address[{\rm7}]{Max Planck Institute for Gravitational Physics (Albert Einstein Institute) and Leibniz Universit\"at Hannover, Germany}

\maketitle \vspace{-3.5mm}{\footnotesize\begin{center} Received September  24, 2015; accepted September 28, 2015
\end{center}}\vspace*{-5mm}

\begin{center}
\rule{16.5cm}{0.4pt}
\parbox{16.5cm}
{\begin{abstract}This paper reviews some of the key enabling technologies for advanced and future laser interferometer gravitational wave detectors, which must combine test masses with the lowest possible optical and acoustic losses, with high stability lasers and various techniques for suppressing noise. Sect. 1 of this paper presents a review of the acoustic properties of test masses.   Sect. 2 reviews the technology of the amorphous dielectric coatings which are currently universally used for the mirrors in advanced laser interferometers, but for which lower acoustic loss would be very advantageous. In sect. 3 a new generation of crystalline optical coatings that offer a substantial reduction in thermal noise is reviewed.  The optical properties of test masses are reviewed in sect. 4, with special focus on the properties of silicon, an important candidate material for future detectors.  Sect. 5 of this paper presents the very low noise, high stability laser technology that underpins all advanced and next generation laser interferometers.
\end{abstract}}
\end{center}\vspace*{-0.6cm}

\begin{center}
\parbox{16.5cm}
{\bf\jiuhao gravitational waves, advanced techniques, thermal noise, coating, laser}
\end{center}

\begin{center}
{}
\Cit{Mitrofanov V P, Chao S, Pan H-W, et al. Technology for the next gravitational wave detectors. Sci China-Phys Mech Astron, 2015, 58: 120404,  doi: 10.1007/s11433-015-5738-8}
\end{center}

\textwidth=178truemm \textheight=236truemm

\wuhao\vspace*{1.5mm}
\tableofcontents
\vspace*{3mm}
\begin{multicols}{2}

\renewcommand{\baselinestretch}{1.08} \baselineskip 12.2pt\parindent=10.8pt

\renewcommand{\thefootnote}


\section{Acoustic properties of test masses for next generation gravitational wave detectors}
\emph{According to the fluctuation-dissipation theorem the test mass thermal noise is determined by its mechanical loss.
In this section we review the main physical mechanisms of loss in perfect and real crystals as well as results of
experimental investigation of mechanical loss in fused silica, sapphire and silicon at temperatures
ranging from 4 K to 300 K.
}

\subsection{Introduction}
Thermal noise of the test masses, their coating and their suspension is considered
as the dominant source of noise in the next generation interferometric gravitational
wave detectors. Regards to the test mass substrate thermal noise the fluctuation-dissipation theorem
predicts that it is determined by acoustic (mechanical) losses of the test mass substrate.
In this section we review the main physical mechanisms of loss in perfect and real crystals as well as results
of experimental investigation of mechanical loss in fused silica, sapphire and silicon
at temperatures ranging from 4 K to 300 K. These materials have small mechanical and
optical losses and can be produced as large-sized samples. Calculation of the power
spectral density of the test mass thermal displacement noise as a function
of mechanical loss is presented in ref. \cite{1Nawrodt}.

\subsection{Mechanisms of losses in mechanical resonators}

The fluctuation-dissipation theorem predicts that thermal noise of the test mass substrate is determined by its acoustic or mechanical losses.
For cyclic loading a complex elastic modulus is introduced. The losses are described as an
imaginary part of an anelastic modulus. In the simplest case, vibration of a mechanical resonator fabricated from an isotropic material can be determined by the Young's modulus of the material:
$Y = Y_1  +{\rm i}Y_2  = Y_1 (1+{\rm i}\: {\rm tg}\phi_m)$, where the loss angle
$\phi_m \approx {\rm tg}\phi_m = Y_2 /Y_1$.
In general, the loss angle is a function of frequency and temperature.
One can quantify the loss angle of a material by measuring the $Q$-factor of a mechanical resonator fabricated from this material. However, it should be borne in mind that the
measured $Q$ is determined by total losses therefore we need to reduce other losses so that they become insignificant. Note also that not always $Q^{-1}$ of a resonator caused by loss in a material is equal to $\phi_m$ of the
material. In some cases it is necessary to take into account mechanical damping dilution and optical dilution factors \cite{2Cagnoli,3Corbitt}.

Losses of a mechanical resonator can be separated into internal and external losses. Internal losses include the bulk loss of the material and the surface loss associated with the surface
layer. External include gas damping from residual gas molecules, the clamping loss, the loss caused by environmental electric and magnetic fields, the loss resulted from back
action of a sensor used for monitoring the resonator vibration and the loss from any other dissipative processes \cite{4Braginsky}.
Considering the bulk losses, we can distinguish those that occur in perfect crystals and can be regarded as fundamental and unremovable. These include the thermoelastic loss and the
phonon-related loss by Akhiezer mechanism. The thermoelastic loss arises from spatially inhomogeneous elastic strains which create the temperature gradients in the vibrating
solid body. They induce heat flows which are accompanied by an entropy increase and a conversion of elastic energy into thermal energy. Zener \cite{5Zener} has developed the analytical model of the
thermoelastic damping for the flexural vibration of an isotropic, homogeneous beam of a thickness $h$:\vspace*{-1mm}
\begin{equation}
Q^{-1}_{\rm te} = \frac{Y \alpha^2 T}{C} \frac{\omega\tau_T}{1+\omega^2 \tau_T^2},\vspace*{-1.5mm}
\label{eqnthermoelasticloss}
\end{equation}
with
\begin {equation}
\tau_T = \frac{C h^2}{\uppi^2 k},
\end {equation}
where $\tau_T$ is the characteristic time for heat transfer across the beam,
$T$ is the absolute temperature, $\alpha$ is the coefficient of thermal expansion,  $C$ is the volumetric heat capacity,
$k $ is the thermal conductivity of the material.
Lifshitz and Roukes \cite{6Lifshitz} obtained a more exact expression for the thermoelastic damping in beams. Calculation of the thermoelastic loss in free-edge circular disks is
presented in refs. \cite{7Li,8Dmitriev}. Thermal fluctuation displacements of the test mass surface as a result of the thermoelastic damping in the test mass substrate are calculated in ref. \cite{9Braginsky2}.
The thermoelastic loss is associated with the thermal expansion of the material which is caused by anharmonicity of a crystal lattice. The anharmonicity leads to another
fundamental loss mechanism arising from interaction between an elastic wave and thermal phonons. Elastic strain changes the phonon frequencies and is different for different
phonon modes thus perturbing the original phonon distribution away from its equilibrium state. The process of restoring thermal equilibrium to the phonon gas is accompanied by
dissipation of the elastic wave energy. The theory of this process was first treated by Akhiezer and was developed further by other authors (see ref.~\cite{10Nowick}). The loss factor $Q^{-1}_{\rm ph}$
of a mechanical resonator determined by Akhiezer damping is given by the expression:
\begin {equation}
Q^{-1}_{\rm ph} = \frac{\sum_i C_i T \left( {\gamma^2_i}-{\overline\gamma}^2\right)}{\rho v^2} \frac{\omega\tau_{\rm ph}}{1+\omega^2 \tau_{\rm ph}^2},
\end {equation}
where $\rho$ is the material density, $v$ is the sound velocity, $\tau_{\rm ph}$ is the mean relaxation time for thermal phonons, $C_i$ is the heat capacity per unit volume of the $\it i$ -th phonon mode in the
crystal, $\gamma_i$ is the Gruneisen parameter for this mode, and $\overline\gamma$  is a quantity obtained by averaging $\gamma_i$ over all directions in the crystal. More information on this
formula applied to the calculation of the loss in sapphire can be found in ref. \cite{4Braginsky}. Comparison of the calculation with the experimental data is also presented.
A modern approach to the analysis of the phonon-phonon loss is given in ref. \cite{11Kunal}. Phonon-electron interaction in crystals also results in additional mechanical loss. This loss is
absent in dielectric materials and is small in semiconductors such as silicon. One can find the calculation of the loss associated with phonon-electron interaction in ref. \cite{12Lindenfeld}.

The above-mentioned dissipation mechanisms set the ultimate lower limit on the loss-factor of mechanical resonators. Real crystals contain various defects such as vacancies,
interstitial and substitutional impurity atoms, dislocations and others. Such defects can act as sources of internal friction. The great majority of internal friction mechanisms may be
regarded as relaxation processes associated with time-dependent transition to another equilibrium state \cite{10Nowick}. When damping is governed by a single relaxation time $\tau$, the
frequency dependence of internal friction is a Debye peak \cite{5Zener}. Its frequency dependence is the same as in eq. (\ref{eqnthermoelasticloss}). The internal friction loss manifests itself as a peak in the temperature dependence of $Q^{-1}$. Examples of their presence in sapphire mechanical resonators are given in ref. \cite{4Braginsky}.

The surface loss is associated with the surface of a vibrating body.  The surface layer can have a structure different from the bulk with damage imparted during fabrication and surface
treatment. Impurities can be incorporated into the surface layer through diffusion of atoms from polishing compounds. Water, atmospheric gases, organic contaminations can be
adsorbed onto the surface layer from the surroundings. When a damaged surface is deformed, relaxation processes associated with change of state of the surface defects can
occur resulting in dissipation.  The best surface quality is achieved as a result of chemical and flame polishing as well as the special treatment of the surface \cite{4Braginsky,13Haucke}. Surface loss effects tend to decrease with decreasing the surface-to-volume ratio for the vibrating body.

Regard to external losses in mechanical resonators we mention here only gas damping and clamping loss as the most important. Other types of external losses in mechanical
resonators have a specific character. The level of the residual gas damping of a mechanical resonator can be calculated using the known formulas for the case of the free
molecular regime and taking into account the squeeze film air damping effect associated with a gap separating a vibrating body and a closely spaced object ~\cite{14Blom,15Bao}.
Suitable vacuum pressure is required in order to make this loss insignificant.
A mounting system of a mechanical resonator can be a source of dissipation due to radiation of the elastic energy into the supporting structure and the interface loss in the contact
area between a vibrating body and a clamp.  This loss is called anchor loss or clamping loss or attachment loss. Generally, the following main methods of reducing these losses
are used: the choice of the resonator design and the vibration mode which provide small clamping loss (e.g. tuning-fork); the clamping in nodal areas of vibration; the use of thin
wire suspension; the use of elements reflecting elastic waves; the use of acoustic insulation layers between the resonator and the clamping structure. At present, numerical
methods are used to calculate the clamping loss \cite{16Frangia}.

\subsection {Results of experimental investigation of acoustic losses in the test mass materials}

When choosing materials for the test masses of next generation gravitational wave detectors we have to take into account several factors. The key factors are small acoustic and
optical losses in these materials as well as the possibility of industrial production of large-sized, high-quality samples of such materials with a weight exceeding 100 kg. Below
we consider three materials: fused silica (SiO$_2$), sapphire (Al$_2$O$_3$) and silicon (Si). The first one is amorphous glass; the other two are single crystals. Although we can not exclude the use
of other materials, for example single crystal calcium fluoride.

\begin{figure}[H]
\centering
\includegraphics[scale=0.98]{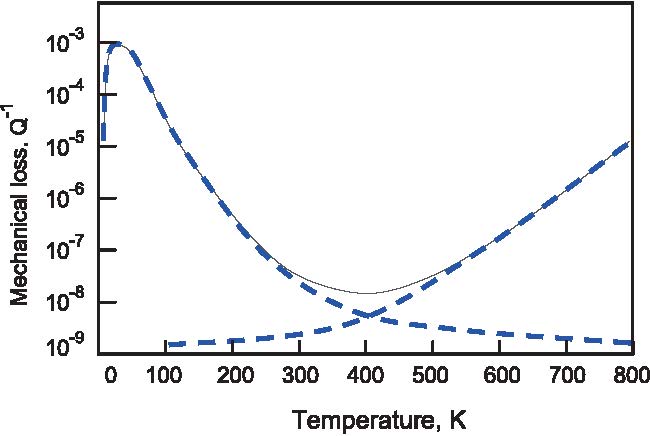}
\caption{(Color online) Characteristic temperature dependence of mechanical losses $Q^{-1}$ in fused silica (see refs.~\cite{17Schnabel,18Lunin}).}
\label{fig:figure1.1}
\end{figure}

Fused silica is a remarkable material widely used due to his excellent mechanical and optical properties, in particular low mechanical and optical losses. The test masses of initial
and advanced interferometric gravitational wave detectors were fabricated from fused silica.  It is a dielectric material with glass structure consisting of a disordered network of
three-dimensional SiO$_2$ tetrahedrons. According to the general idea of the influence of impurities on the dissipation the minimal loss is achieved in very pure silica. The main
dissipation mechanism in fused silica mechanical resonators is associated with structure relaxation when the oxygen atom in the Si---O---Si bonds transfers between two
equilibrium states under the action of mechanical stresses. This process modeled as a two-level tunneling system results in a wide relaxation peak of loss.  At audio frequencies
the low-temperature peak with a level of peak loss of about $10^{-3}$ is observed in the temperature range 20--50 K \cite{17Schnabel}.  At elevated temperatures there is another loss peak. Tails of these peaks determine loss of fused silica resonators obtained at room temperatures \cite{18Lunin} (see Figure~\ref{fig:figure1.1}).

Results of multiple measurements of $Q$-factors of fused silica
mechanical resonators obtained in~ different~ labs~ were

\noindent collected for generating the empirical model of the room temperature loss in fused silica. The loss function takes the form
including terms with coefficients $C_1, C_2$ and $C_3$ describing the surface loss and the frequency dependent bulk loss \cite{19Penn}:
\begin {equation}
\phi = C_1 \left(\frac V S\right)^{-1} + C_2 (f/1 Hz)^{C_3},
\end {equation}
where $(V/S)$ is the volume-to-surface ratio given in mm. The results were presented for two types of pure fused silica: Suprasil 2 and Suprasil 312. The surface of the samples
was polished using the superpolishing technique and flame polishing for thin fused silica fibers. The following coefficients were found for Suprasil 312: $C_1 = 6.5$ pm, $C_2
= 7.6\times 10^{-12}$, $C_3 = 0.77$ \cite{19Penn}.
Fused silica has a low level of the thermoelastic loss at room temperatures due to the relatively small thermal expansion coefficient ($\alpha = 0.5\times 10^{-6}\ {\rm K}^{-1}$). Another
advantage of fused silica as an amorphous material is the possibility of bonding its pieces into the monolithic construction with the help of welding as well as pulling silica fibers
with a diameter from a few to hundreds of microns \cite{20Heptonstall}. This allowed creation of monolithic fused silica pendulum with relaxation time of about 3 years which corresponds to
the quality factor $Q = 1\times 10^8$ \cite{21Braginsky3}. The disadvantage of fused silica is its low thermal conductivity ($k = 1.2\:{\rm  Wm}^{-1}{\rm K}^{-1}$ at 300 K) which prevents an increase of the
optical power in the gravitational wave interferometer due to thermal distortions of the test masses � mirrors. As already mentioned the high level of low temperature mechanical loss of
fused silica does not allow using it in the cryogenic detectors.

\begin{figure}[H]
\centering
\includegraphics[scale=1]{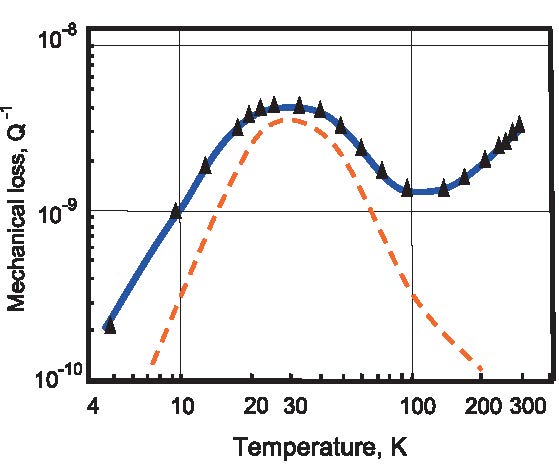}\vspace{-2mm}
\caption{(Color online) Calculated (dashed line) and measured (solid curve with triangles) temperature dependencies of mechanical loss $Q^{-1}$ in sapphire resonator with natural frequency of
38 kHz (see ref.~\cite{4Braginsky}).}
\label{fi2:temp}
\end{figure}

Sapphire is a crystalline form of corundum $\alpha$-Al$_2$O$_3$ belonging to the trigonal crystal system. It is a hard material with a high melting point of $2030^{\circ}$C. It has a
high chemical resistance and can be etched at high temperatures with a restricted set of chemical agents. Sapphire has one of the highest Debye temperatures $T_{\rm D}$ = 1947 K.
Sapphire monocrystals have low dislocation mobility and accordingly low level of internal friction. The temperature dependence of the loss $Q^{-1}$ measured in the sapphire
mechanical resonator with the natural frequency of 38 kHz is shown in Figure~\ref{fi2:temp} \cite{4Braginsky}.

The resonator was made from sapphire in the form of a cylinder with a length of 137 mm and a
diameter of 44 mm. The axis of the cylinder was directed at an angle of $60 ^{\circ}$ to the optical axis of the crystal.  The cylinder was suspended with a silk thread loop
embracing it in its middle. The resonant vibration was excited on the fundamental longitudinal mode of the resonator. One can see a peak of loss at about 30 K, the same
temperature as the known peak in the thermal conductivity of sapphire. Since the loss caused by the phonon-phonon Akhiezer mechanism is proportional to the relaxation time of
thermal phonons and thereby to the thermal conductivity, the observed peak of loss can be associated with phonon-phonon interaction. There is a good coincidence between the
calculated and the measured losses at temperatures near the peak of loss. Other loss mechanisms apparently the surface loss and the clamping loss prevail in other temperature ranges. The
thermoelastic loss is negligible for this longitudinal vibrational mode. The record $Q = 3\times 10^8$ at 300 K and $Q = 5\times 10^9$ at 5 K were obtained for the sapphire
mechanical resonator. Nevertheless sapphire is not used as a material for the test mass of the room temperature gravitational wave detectors because of its high thermoelastic loss
associated with the relatively large thermal expansion coefficient ($\alpha = 6.6 \times 10^{-6}~ {\rm K}^{-1}$). It significantly decreases with temperature so the sapphire test masses are used in
the cryogenic detectors such as KAGRA in Japan \cite{22Hirose}. The level of the test mass thermal noise which will be obtained in the cryogenic detector based on sapphire test masses
depends on the quality of sapphire fibers used for their suspension and the perfection of the technique used for bonding of the fibers to the test mass.

\begin{figure}[H]
\centering
\includegraphics[scale=1]{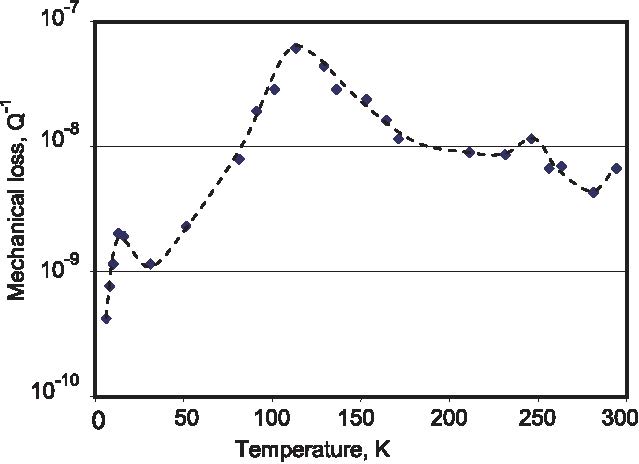}
\caption{(Color online) Temperature dependence of mechanical loss $Q^{-1}$ in silicon resonator with natural frequency of 7420 Hz (see ref.~\cite{24Mitrofanov}).}
\label{fig3:Temp2}
\end{figure}

Single crystalline silicon is known as an excellent material for fabrication of micro- and nanomechanical high $Q$ resonators which are used in a vast number of different devices
although the best $Q$-factors were obtain in large-sized mechanical resonators due to small influence of the surface loss in such resonators \cite{23McGuigan}. Figure~\ref{fig3:Temp2}  shows the~ temperature~
de-

\vspace{2mm}

\vspace{4mm}
\noindent pendence of the loss-factor $Q^{-1}$ of a silicon resonator excited on the fundamental longitudinal mode with a resonant frequency of 7420 Hz \cite{24Mitrofanov}. The silicon resonator was a
cylinder with a diameter of 77.5 mm and a length of 604 mm. Its mass was about 6 kg.

The cylinder axis was parallel to the [111] direction in the crystal.  Czochralski grown
silicon was weakly doped with phosphorus and had a resistivity of about  1 kOhm$\cdot$cm at room temperature. The ends surfaces and the barrel of the cylinder were polished
with diamond pastes in order to reduce the surface loss. The resonator was suspended by a loop of 0.3 mm-diameter polished molybdenum wire and was placed into a vacuum
chamber inside a liquid helium cryostat. Two peaks at $T$ $\approx$ 13 K and $T$ $\approx$ 115 K were observed in the temperature dependence of losses in the silicon resonator. They are similar to the
peaks of losses observed in ref. \cite{23McGuigan} where boron-doped silicon resonator with a resistivity of about $ 4\: {\rm Ohm}\cdot {\rm cm}$ was investigated and where a 13 K peak was
considerably higher. This allows assuming that this peak is associated with dopant impurity atoms in the silicon crystal. The loss peak similar to the 115 K peak was observed in ref.
\cite{23McGuigan}. Note that temperature of the peak shifts slightly when changing the natural frequency of the resonator in accordance with the theory of relaxation process. Several possible
loss mechanisms which can cause the peak was assumed but so far there is no reliable evidence to support one or the other of the proposed mechanisms.
It is interesting that both peaks are in the temperature ranges where the thermal expansion coefficient of silicon crosses zero at $T$ $\approx$ 18 K and 124 K. At these temperatures,
there is no thermoelastic loss in the test masses and correspondingly thermoelastic noise is absent.  Silicon is not used as the test mass material at the
room temperature because it has the relatively large thermal expansion coefficient ($\alpha = 2.6 \times 10^{-6}~ {\rm K}^{-1}$). Silicon is considered as a prospective material for the test masses working at the temperature of zero
thermal expansion coefficient. Suppression of the thermoelastic loss in silicon at $\approx$ 124 K was
demonstrated in refs. \cite{26Reid,27Nawrodt2,28Prokhorov}. However, it remains an open question as to what minimal level of mechanical loss can be achieved for the silicon test mass in this case as well as what minimal loss can be obtained in the suspension ribbons/fibers and in their bonding to the test mass. It is
important for calculation of thermal noise of the test masses and their suspensions in the future gravitational wave detectors.


\section{Amorphous coatings for laser interference gravitational wave detectors}
\emph{Coating thermal noise of the mirror is the dominant noise source for laser interference gravitational wave detector at its most sensitive frequency range about 100 Hz. Coatings of the mirror is composed of pairs of alternating high and low refractive index thin films with thickness of quarter-wavelength. Currently, the materials used in the coatings are silica and titanium-doped tantala in amorphous form, and the deposition method is ion beam sputtering. In this section, we review the amorphous coating development for the laser interference gravitational wave detector with respect to the coating layer structure, coating methods, and coating materials. Conventional and potential nonconventional layer structures are reviewed. Ion beam sputter deposition method as well as chemical vapor deposition method is introduced. Potential high and low refractive index amorphous coating materials are reviewed with particular focus on the room temperature and cryogenic temperature mechanical loss and the optical loss.}

\subsection{Dielectric multi-layer stack}\label{sec:Di}

\subsubsection{General}
Mirrors for the laser interference gravitational wave detector consist of a dielectric multi-layer stack with pairs of alternating high and low refractive index thin films each has one-quarter wavelength (QW) optical thickness, i.e. physical thickness times the refractive index, deposited on a substrate. Reflectance of the mirror for normal angle of incidence in free space and with optical isotropic materials is given as \cite{1_Macleod}:
\begin{equation}\label{eq1}
R=\left[\frac{1-(n_{\rm H}/n_{\rm L})^{2p}(n_{\rm H}^{2}/n_{\rm S})}{1+(n_{\rm H}/n_{\rm L})^{2p}(n_{\rm H}^{2}/n_{\rm S})}\right]^{2},
\end{equation}

\noindent where $n_{\rm H}, n_{\rm L}$ and $n_{\rm S}$ are the refractive indices of the high index layer (H), low index layer (L) and the substrate, respectively, and $p$ is the number of pairs.
Usually, the coatings are designed such that there are odd number layers with the high index layer outermost and innermost. An optically inert layer with one-half optical thickness could be added on top of the stack for protection purpose if necessary. Transmittance of the mirror decreases by a factor of $(n_{\rm L}/n_{\rm H})^{2}$ with addition of one extra pair if there is no optical loss.

\textit{Brownian thermal noise} of the coatings is the limiting factor for detection sensitivity of the laser interference gravitational wave detector in the frequency range around 100 Hz \cite{2_LIGO1}. Thermal noise of a system is related to the mechanical loss of the system through the fluctuation-dissipation theorem \cite{3_Callen}. Power spectrum of the thermal noise of the coatings is proportional to the mechanical loss, characterized by the loss angle, and the thickness of the coating materials \cite{4_Harry1}. Compendium of mirror thermal noise was well categorized \cite{5_Harry2}. Among the various sources of thermal noises for the coatings, Brownian and thermo-optic (combination of thermoelastic and thermorefractive) noises dominate. However, fluctuation of optical path length from thermoelastic and thermorefractive are opposite, i.e. change of physical path length due to thermal expansion from temperature fluctuation is opposite to the change of refractive index from temperature fluctuation, and can be suppressed coherently \cite{6_Gorodetsky,7_Harry3} leaving Brownian thermal noise the major noise for the coatings. Advanced LIGO and VIRGO are operating in room temperature, and KAGRA \cite{8_Aso} is operating in cryogenic temperature, Voyager \cite{2_LIGO1} and ET-LF \cite{9_Abernathy} are proposed to operate in cryogenic temperature. Reduction of mechanical loss hence the Brownian thermal noise for the coating materials in the room temperature for the existing detectors and in the cryogenic temperature for the next generation detectors are major focus of the R\&D works on coatings development.

\textit{Optical loss} in the coatings, i.e. scatter and absorption loss, limits the increment of reflectance from adding more pairs and they usually are the major sources for limiting the quality factor ($Q$) of the optical resonator. Temperature variation in the coatings due to absorption and scattering can cause serious mirror damage in high power application \cite{10_Stolz} and phase distortion in precision measurement \cite{11_Harry4}. Optical absorption due to oxygen deficiency in the amorphous heavy metal oxides high index material such as Ta$_2$O$_5$ and TiO$_2$, which are the most commonly used high index materials for the visible and near IR wavelength range, is the major source for optical absorption asides from other sources such as impurity absorption. Prolonged high temperature annealing in ambient after deposition for oxygen to diffuse in the layers and fill the oxygen vacancies is a common practice for coating vendor to reduce the optical absorption of their product \cite{12_Gibson}. Thermal annealing was reported to also reduce the mechanical loss for many thin film materials \cite{13_MartinIW,14_Penn,15_Abernathy2}. However, the annealing temperature must not exceed the crystallization temperature for the amorphous film to become polycrystalline for which the grain boundaries and interface roughness will increase the optical scattering \cite{16_MartinN} and furthermore, the mechanical loss will likely to increase due to friction between the grain boundaries and in the rough interfaces.

Eq.~(\ref{eq1}) implies that it is of great advantage to use high and low index materials with as large difference in index as possible. Firstly, the number of pairs can be reduced and yet the same level of reflectance maintained, and secondly, the physical thickness of the layer can be reduced for QW optical thickness with higher index. The overall effect is reduction in mechanical loss and thus thermal noise by reduction in total thickness of the multi-layer stack.

\subsubsection{Nonconventional layer structure}
The standing wave distribution in the QW stack is that the power peaks located at the HL layer interfaces and the peak height reduces rapidly from the coating front, i.e. facing the incoming laser beam, toward the coating end \cite{17_Chao1}. Most of the coating absorption and scattering, therefore, occur in the first few pairs and at their interfaces. This lead to the idea of separating the front few pairs of the coatings from the rest pairs and the bulk substrate by free space or a spacer with low thermal noise and with separation of odd integer times of quarter-wavelength, i.e. anti-resonance, such that thermal noise from the main mass of the mirror contribute little to the reflected phase fluctuation and yet retain the reflectivity \cite{18_Khalili}, this kind of structure in known as \textit{Khalili cavity}. Similar idea was applied to use coating material that has low optical absorption but high mechanical loss in the front few pairs and coating material that has low mechanical loss but inferior optical absorption in the back pairs to reduce the mechanical loss and yet limiting the optical absorption of the whole mirror to the low level \cite{19_Steinlechner}.

There are advantages to design the layer thickness away from the conventional quarter wave in view of reducing the mechanical loss of the coatings. When the mechanical loss of the high and the low index thin film materials differs significantly, it was proposed \cite{20_Harry5} that the layer thicknesses and number of pairs could be optimized via the formulations of the reflectance and the loss angle of the multi-layer stack, such that thickness of the high mechanical loss material be reduced and that of the low mechanical loss material be increased to reduce the overall mechanical loss of the stack and yet to maintain the desired reflectance. It was also proposed \cite{21_Pinto} according to the effective medium analysis that replacing the QW layer by a stack of alternating HL layers each has thickness in the scale of nanometer, referred to as the \textit{nano-layer} structure, could reduce the mechanical loss, and it was later discovered experimentally \cite{22_PanHW} that the nano-layer structure can sustain higher temperature of annealing than the single QW layer without crystallization. Furthermore, nano-layer structures with equal total thickness but with thinner layers, i.e. more number of layers, can sustain higher temperature of annealing than those with thicker layers, hence the mechanical loss can be reduced further by higher temperature annealing \cite{23_Chao2}.

\subsection{Deposition methods}

\subsubsection{Ion beam sputter deposition (IBSD)}
There are various existing methods for depositing amorphous optical thin films in the industry \cite{24_Harry6}. Conventional methods for depositing optical thin films are evaporation type in which the material to be deposited is heated by resistance heating or electron beam bombardment and the evaporant condense on the substrate to form the film. Also there is glow discharge sputter type in which the substrate and the sputter target constitute part of the glow discharge system in DC, RF or magnetron configurations to sputter off the material from the target and condense on the substrate. It was invented in the mid-70s \cite{25_Wei1}, driven by the need of high quality mirrors for the ring laser gyroscope, to use ion beam to bombard the target to sputter off the material and condense to form the thin film. And it was discovered \cite{26_Wei2} that the optical quality of the amorphous thin film coated by ion beam sputter deposition (IBSD) are superior to that of the film coated by other methods in terms of higher refractive index, lower optical loss and better mechanical strength. It was then realized \cite{27_Maissel} that the good qualities for the IBSD thin films are attributed to the higher kinetic energy of the atoms impinging the substrate prior to condensation, which is in favor of the subsequent thin film growth for a more densely packed amorphous structure. Nowadays, IBSD is the standard coating method for high-end optical coating applications. Disadvantage for IBSD is its low throughput. The deposition rate is low such that it takes hours to deposit a multi-layer dielectric high reflector. The simplest ion beam source is the Kaufman type in which a hot filament is used in the plasma chamber, usually with argon plasma, and the heavy ion plasma be extracted to generate the high energy beam for sputtering. Advanced ion beam source adopts the more efficient electron cyclotron resonance (ECR) plasma generation method in which a microwave of 2.45 GHz is sent into the discharge chamber where a static magnetic field with strength of 856 Gauss is set to create ECR and produce plasma more efficiently. The ERC method is cleaner in which there is no contamination from the filament.

The size of the mirror for laser interference gravitational wave detector is typically $\sim40$ cm in diameter with curvature. It is ultimate important to deposit the multi-layer stack with high degree uniformity across the entire surface and accuracy for thickness control in the IBSD process. The thickness is controlled in two ways, first by using the quartz crystal monitor and second by using the optical monitor. Light of the design wavelength is shined on the witness piece aside the substrate, the reflectance or the transmittance of the witness piece undergoes extrema when the thickness of the film reaches integer multiplication of QW, and the deposition of the layer is terminated at the extrema. Both monitoring method needs calibration with the true thickness on the substrate. The angular distribution of the sputtered atom is usually under-cosine \cite{28_Wehner}, and in order to compensate for the non-uniform distribution of the incoming atoms, the substrates are rotated in a planetary fashion, and a mask with proper shape is designed to shade the rotating substrate unevenly to yield a uniform coating. Other geometrical configurations in the coating chamber such as substrate tilt, ion beam direction, distances between elements need to be fine-tuned in order to optimize the uniformity. Coating uniformity with amplitude smaller than 0.5 nm for all the Zernike terms, i.e. polynomial components of the surface contour for a circular substrate, was achieved for coating the mirrors of Advanced LIGO \cite{29_Pinard1}. Reports for low loss coatings for LIGO and VIRGO mirrors using IBSD can also be found in refs. \cite{30_Pinard2,31_Beauville,32_Netterfiled}.

\subsubsection{Chemical vapor deposition (CVD)}
It is advantageous to explore methods other than IBSD for depositing the amorphous coatings for the mirrors of the laser interference gravitational wave detector. Chemical vapor deposition (CVD) is a process where by sending reaction gases into the chamber, sequence of chemical reactions occurs between the reaction gases and/or between the gas and the substrate and the product condenses on the substrate to form the film. Various techniques were used to enhance the reaction rate; low pressure CVD (LPCVD) process allows higher substrate temperature to expedite the reaction rate and plasma enhanced CVD (PECVD) using RF generator to create plasma of the reaction substances to enhance the reaction rate. Other means of CVD, e.g. Metal-organic CVD (MOCVD) and molecular-beam-epitaxial (MBE), can be used to fabricate high quality epitaxial crystalline film on the substrate and they are introduced elsewhere in this special reports.

Silicon, for its low cryogenic mechanical loss \cite{26Reid} and existing mature infrastructure for manufacturing high quality and large scale wafer in the semiconductor IC industry, is proposed to be used as the mirror substrate for the next generation laser interference gravitational wave detector operated at 1550 nm wavelength \cite{2_LIGO1,9_Abernathy}. Silicon wafer with diameter of $16''$ and $18''$, equal to and larger than the mirror of the gravitational wave detector, is nowadays used for the IC manufacturing. Various CVD methods are commonly used in the in-line silicon IC manufacturing process to deposit thin films on the silicon wafer. It would be advantageous for the next generation detector if some of the IC thin film materials can be used for the mirror coatings; then the silicon IC manufacture process can serve the mirror fabrication and the IC manufacturers could become the mirror providers. There are some potential amorphous materials used in the silicon IC and deposited by CVD processes that could be explored for the purpose. Among these are amorphous silicon (a-Si, refractive index 3.4), SiO$_2$ (refractive index 1.46) and silicon nitride (SiN$_x$). Refractive index of SiN$_x$ varies with concentration and process condition, and it is within the range of 1.8--2.6 \cite{34_Stoffel}. Therefore, SiN$_x$ can serve either as high or low index layer in the multi-layer stack. Various H/L pair combinations are possible: a-Si/SiO$_2$, SiN$_x$/SiO$_2$, a-Si/SiN$_x$, and SiN$_x$(H)/SiN$_{x}$(L). These materials are fabricated by using the CVD process and readily in amorphous form \cite{34_Stoffel,35_Nguyen}. The major focus of research on these materials will be mechanical loss, both in room and cryogenic temperature, and optical losses, see the following section.

\subsection{Amorphous thin film materials}
Ring laser gyroscope (RLG)~ and ~high~ power ~laser ~systems

\noindent have set a high level standard on the optical qualities for the dielectric multilayer mirror \cite{10_Stolz,36_Kalb}. It provides a good starting for materials selection for the coatings of the laser interference gravitational wave detector. Conventional coating materials for RLG and many high power laser systems in the visible and near IR includes amorphous Ta$_2$O$_5$ (tantala) as the high index material and SiO$_2$ (silica) as the low index material, both are deposited by IBSD and the coatings undergoes the post deposition heat treatment to reduce the optical loss. Silica and Ti-doped tantala for the low and high index layers, respectively, are currently used for the Advanced LIGO and VIRGO. Mechanical loss and optical properties, both in room temperature and cryogenic temperature, of these materials and their related materials are the focus of R$\&$D for the gravitational wave detector. Lower mechanical and optical losses are primary goals for investigation and new material exploration. A good summary of the potential materials was given in 2009 \cite{37_France}, and here we shall emphasize on the new developments afterwards.

\subsubsection{Low index material-IBSD silica}
There are many low index thin film materials \cite{24_Harry6}, e.g. various oxide, fluoride, and chalcogenide, have been used in optical applications for the UV, visible and IR wavelength range. Some of these materials suffer from low crystallization temperature, high mechanical loss, hydrophilic, poor adhesion, or high optical loss and are disqualified to be the candidate for the purpose. IBSD silica is currently the one if not the only one amorphous low index material under investigation for the coatings of the laser interference gravitation wave detector. More efforts should be invested on exploring new suitable low index amorphous material other than silica. IBSD silica film has the room temperature mechanical loss in the lower-$10^{-4}$ range for frequency of several hundreds to three thousands Hertz range \cite{38_MartinI,39_Penn2}. Thermal annealing reduced the room temperature mechanical loss down to the lower-$10^{-5}$ range \cite{39_Penn2,40_Penn3}. There is approximately one order of magnitude reduction in the room temperature mechanical loss by thermal annealing for IBSD silica film.

IBSD silica film has a clearly identified loss peak around 20 K \cite{41_MartinIW2} and the peak height reached upper-$10^{-4}$ and the activation energy was estimated to be 32 meV. This low temperature peak is consistent with the peak appeared in bulk silica although with small differences in peak position and activation energy. The cryogenic loss peak of bulk silica was attributed \cite{42_Gilroy} to Si---O bond, e.g. bond angle or bond length, reorientation between two asymmetrical configurations via an asymmetrical two level system (TLS) potential barrier in the amorphous state. Loss angle can be modeled for the TLS with given distribution function of the barrier height \cite{43_Topp}. This theory is the basis for analysis on the mechanical loss of the IBSD amorphous films by many authors \cite{44_MartinI2,45_Murray}.

The cryogenic loss peak implies a drawback for using silica film in the coatings for detector operating in cryogenic temperature. Since silica is currently the only low index material that has been studied intensively for mirror coatings of gravitational wave detector, it is therefore important to explore other low index materials for next generation detector. Silica-based low index materials would be worth investigating, for example, it would be interesting to investigate the effect of low level doping for silica to modify its cryogenic properties and yet not to increase the index significantly and maintain its amorphous form. Although it is generally considered that the cryogenics loss peak around 20 K and the high loss plateau below 10 K is universal to amorphous materials due to the TLS \cite{46_Phillips, 47_Anderson}, however, doping of 1\% H in a-Si for example was shown to reduce the cryogenic loss significantly \cite{48_Liu1}. On the other hand, evidence showed that increase in local mass density and structural order through higher temperature deposition of the amorphous film also removed TLS and hence reduced the cryogenic loss \cite{49_Liu2, 50_Liu3}. Nano-layer structure of TiO$_2$/SiO$_2$ that has an equivalent low index, i.e. silica dominant, is also worth exploring in view of that pure IBSD TiO$_2$ might not have cryogenic peak \cite{51_Pinto3}. It would be therefore valuable to investigate the cryogenic behavior of the TiO$_2$/SiO$_2$ nano-layer structure.

\subsubsection{High index materials-tantala and Ti-doped tantala}
Annealed IBSD tantala (Ta$_2$O$_5$) film has room temperature loss angle at the mid-$10^{-4}$ range \cite{13_MartinIW, 40_Penn3}, more than one order of magnitude larger than that of the silica film and is the dominant room temperature mechanical loss for Ta$_2$O$_5$/SiO$_2$ coatings. There is cryogenic loss peak for tantala near 20 K as well, and heat treatment increases the cryogenic loss below 200 K \cite{13_MartinIW} that is in contrast to the annealing effect of many other materials. The peak value of the loss for $600  ^\circ$C annealed tantala was nearly $\sim1\times10^{-3}$. Since $600 ^\circ$C annealing is necessary for reducing the optical loss of the Ta$_2$O$_5$/SiO$_2$ QW coatings, therefore, the high cryogenic mechanical loss for both Ta$_2$O$_5$ and SiO$_2$ makes the conventional coatings unsuitable for the laser interference gravitational wave detector operating at the cryogenic temperature.

It was discovered \cite{52_Harry7} that the room temperature loss angles of tantala were reduced by nearly half by doping the tantala with 20\%--30\% titania (TiO$_2$) and with post-deposition annealing at $600 ^\circ$C. The $14.5\%$ Ti-doped tantala film was found to have a cryogenic loss peak near 20 K, same as the un-doped tantala film; activation energy of the $14.5\%$ Ti-doped and un-doped films were 42 meV \cite{44_MartinI2} and 28.6 meV \cite{53_MartinIW3} respectively. However, the peak height was reduced and the peak width was increased for the doped film, which was shown to be the consequence of TiO$_2$ doping by modeling with the TLS analysis \cite{53_MartinIW3}. $55\%$ Ti-doping with $600 ^\circ$C anneal further suppressed the cryogenic loss peak \cite{54_MartinI3}. In the same report, $75\%$ Ti-doping with $600 ^{\circ}$C anneal showed a very low and broad loss peak, the peak height was at upper-$10^{-5}$, but the no-anneal and the $400 ^{\circ}$C-annealed samples showed less-profound peak at the upper-$10^{-4}$ level. $75\%$ Ti-doping with $600 ^\circ$C anneal is currently the best found Ti-doped tantala material in terms of mechanical loss; about one order of magnitude superior to tantala and to other Ti-doped tantala. Accompanied with the $75\%$-Ti doping and $600 ^\circ$C anneal however was the appearance of crystallization, but no report on the kind of crystal. Pure amorphous IBSD TiO$_2$ film was known to crystallize into anatase crystal form at $200 ^{\circ}$C annealing \cite{55_Chao3}, therefore, it was likely that the $75\%$ Ti-doped tantala might crystallize into anatase TiO$_2$ upon annealing. Optical scattering measurement on $75\%$ Ti-doped tantala film will be helpful to justify the usage of this material. Preliminary result \cite{51_Pinto3} indicated that pure IBSD TiO$_2$ film does not have the cryogenic peak, therefore, it is likely that TiO$_2$-doping might help to reduce the cryogenic peak of the tantala in addition to maintaining the high index for the mixed film. It will be of great value to seek optimization between the Ti-doping and heat treatment for the tantala film for low cryogenic mechanical loss and optical loss.

\subsubsection{Other high index oxides}
\textit{Hafnia} (HfO$_2$) deposited by IBSD was reported \cite{15_Abernathy2} to have cryogenic mechanical loss in the lower-$10^{-4}$ range below 100 K, an order of magnitude lower than that of tantala and Ti-doped tantala. The cryogenic peak was vague. The loss remained approximately same level when annealed up to $400 ^\circ$C regardless of the appearance of crystallization. $30\%$ silica-doped hafnia inhibited the crystallization and showed $\sim4.5\times10^{-4}$ and $3\times10^{-4}$ loss angle upon $600 ^\circ$C and $400 ^\circ$C annealing, respectively \cite{56_Murry3, 57_Penn4}. The material seems promising in terms of loss angle but it will be important to obtain the optical absorption of the material and estimate the thermal noise power spectrum for the coatings.

\textit{Zirconia} (ZrO$_2$) deposited by IBSD, with refractive index around 2.0, has been investigated \cite{58_Flaminio}. The room temperature mechanical loss of IBSD zirconia was found to be $2.3\times10^{-4}$; it about ten times lower than that of tantala. However, extremely large stress and high optical absorption made it doubtful for the purpose. Attempts to reduce the stress by doping Ti and W did bring down the stress but not the optical absorption. Efforts of doping the zirconia with silica or tantala were carried out \cite{57_Penn4} in order to increase the crystallization temperature of zirconia such that higher temperature anneal could be performed to reduce the optical absorption and yet to maintain or even reduce the mechanical loss, the effects are remained to be seen.

\textit{Niobium pentaoxide} (Nb$_2$O$_5$) deposited by IBSD, with refractive index 2.21, was reported \cite{58_Flaminio} to have room temperature loss angle of $4.6\times10^{-4}$ but with optical absorption twice as higher than that of tantala, and no further investigation was reported.

\textit{TiO$_2$/SiO$_2$ nano-layer and TiO$_2$-SiO$_2$ mixed film} are potential high index materials, although they can also serve as low index materials with larger proportion of SiO$_2$. Room temperature mechanical-loss of the TiO$_2$/SiO$_2$ nano-layer composites were shown to be reduced by thermal annealing \cite{23_Chao2,59_Deslvo}. Nano-layers with equal total thickness but with more number of TiO$_2$/SiO$_2$ pair, i.e. thinner for each layer, were shown to sustain higher temperature of annealing without crystallization \cite{22_PanHW}. Room temperature loss angle of the nano-layer structure was found to be reduced down to the lower-$10^{-4}$ range after annealing and the films remained amorphous form \cite{23_Chao2}, it is therefore a potential candidate for improving coatings for room temperature gravitational wave detector. Cryogenic loss measurement will be important to justify the usage of the nano-layer structure for cryogenic detector in view of that preliminary result showed that there was no cryogenic peak for the pure IBSD TiO$_2$ film \cite{51_Pinto3}. On the other hand, IBSD TiO$_2$-SiO$_2$ mixed films was reported to have large range of refractive index and annealing temperature tuning capability with good optical quality \cite{55_Chao3}. Room temperature mechanical loss of the IBSD TiO$_2$-SiO$_2$ mixed film was slightly higher than that of Ti-doped tantala \cite{60_Netterfield2}. It will also be valuable to investigate the cryogenic mechanical loss of TiO$_2$-SiO$_2$ mixed film.

\subsubsection{Amorphous silicon}
Amorphous silicon (a-Si) has refractive index as high as 3.4 at 1550 nm. As pointed out previously, less number of HL pairs is required and hence thinner stack by using high index film in the QW stack to yield lower mechanical loss. IBSD a-Si film was reported to have cryogenic loss in the range of mid-$10^{-5}$ to lower-$10^{-4}$ and thermal anneal up to $450^\circ$C reduced the cryogenic loss down to lower-$10^{-5}$ range \cite{45_Murray}. In the same report, bilayer of IBSD a-Si/SiO$_2$ was demonstrated to have cryogenic loss about mid-$10^{-4}$, and Brownian thermal noise at 20 K for mirror coatings consisted of a-Si/SiO$_2$ pairs was estimated to be $55\%$ lower than that of the Advanced LIGO specification. A-Si outperforms tantala and Ti-doped tantala in terms of mechanical loss as the high index materials. However, optical absorption of the IBSD a-Si is high although it was reported that $450 ^\circ$C thermal annealing reduced the optical absorption of IBSD a-Si down to 3800 ppm \cite{61_MartinI4}. An alternative layer structure with multi-material coatings as pointed out previously was purposed \cite{19_Steinlechner} by using conventional tantala/silica for the front pairs in the QW stack and a-Si/SiO$_2$ for the back pairs, the overall Brownian thermal noise was reduced $25\%$ and the optical absorption was at level of 5 ppm.

\subsubsection{Silicon nitride}
Silicon nitride films are readily deposited into amorphous form by using CVD methods such as LPCVD and PECVD. Composition, refractive index, optical absorption and elastic properties are highly process dependent \cite{34_Stoffel}. Composition of the film varies from silicon-rich to nitrogen-rich by varying the process parameters such as gas flow rate and substrate temperature, therefore, the film is designated as SiN$_x$ in general. The refractive index varies in a wide range between 1.8 and 2.6 \cite{34_Stoffel} depending on the process and the composition. The optical extinction coefficient are smaller than $10^{-3}$ \cite{62_Philip,63_Poenar,64_Chao5} at near IR wavelength. The SiN$_x$ films in general are stressful. The commercial SiN$_x$ membranes with stress, in analogy to a taut drum head, for use as sample holder for transmission electron microscope, was reported to have mechanical $Q$ exceeded $10^6$ at room temperature and exceeded $10^7$ at 0.3 K for frequency of 133 kHz, and the extinction coefficient was $1.6\times10^{-4}$ \cite{65_Zwick}. Similar membrane SiN$_x$ was reported to have mechanical dissipation ($Q^{-1}$) ranged from mid-$10^{-7}$ to lower-$10^{-6}$ in the temperature range from 3 K to room temperature at 1.5 MHz, three orders of magnitude lower than that of silica, and the room temperature dissipation was lower than that of crystalline silicon \cite{66_Southworth}. In contrast to the very low mechanical loss of the stressed SiN$_x$ membrance, the stress-relieved SiN$_x$ membrance, i.e. in cantilever shape where three sides were unsupported, had dissipation nearly two orders of magnitude higher than that of the stressed SiN$_x$ membrane but still about a factor of two lower than that of silica \cite{66_Southworth}. Theoretical modeling for the low mechanical loss of the stressed SiN$_x$ was reported \cite{67_Wu}, and it was shown that the stress induced reduction of dissipation can be attributed to increasing of the barrier height of the TLS or decreasing of the coupling between phonons and the TLS by the stress. It was also shown that dissipation dilution, where stiffness of the SiN$_x$ is increased by the external stress to increase the $Q$ but without increasing the loss, did not give a magnitude that matched the experimental observation.

Room temperature mechanical losses of PECVD SiN$_x$ with $x$ ranging from 0.40 to 0.87 were studied within the frequency range of interest for gravitational wave detector \cite{64_Chao5, 69_Chao6}. The films were deposited on a silicon cantilever. Stresses of the films were tensile and increased from 120 MPa to 413 MPa with $x$ increased from 0.40 to 0.87. For 413 MPa high stress film, $Q$ of the film+sub cantilever bending modes were larger than that for the uncoated substrate, referred to as phenomenon of inversion, but none of the torsion modes showed the inversion [50]. The frequency changes were too small to count for dissipation dilution \cite{68_Juang}. Since torsion modes of the silicon cantilever are not susceptible to thermoelastic loss, therefore, a possible mechanism for the inversion is that the thermoelastic loss of the silicon substrate might be reduced under the stress. For 120 MPa low stress SiN$_x$ film, where there was no inversion for the low frequency bending mode, the room temperature loss angle of the film was found to be in the upper-$10^{-5}$ range at $107$ Hz \cite{68_Juang}. A double-side coating and data reduction method were proposed \cite{69_Chao6} to eliminate the phenomenon of inversion for the high stress film. By using this method, the room temperature loss angles of the high stress SiN$_{0.87}$ film was found to be in the lower-$10^{-5}$ range at $107$ Hz, a value that is much lower than that of the Ti-doped tantala.

\subsection{Correlations between atomic structure and mechanical loss}

Efforts have been pursuit to understand the relationship between the atomic structure and the mechanical loss of the amorphous materials that have undergone heat treatment or doping. Exploring the near neighbor distances and distributions of IBSD tantala and Ti-doped tantala by using the reduced density function (RDF) that was obtained by Fourier transforming the electron diffraction pattern of the samples resulted in a clear correlation between the decreasing in mechanical loss and increasing in homogeneity of the first and the second nearest neighbor distance as the Ti-doping concentration increased \cite{70_Bassiri}. Recent report \cite{50_Liu3, 49_Liu2} on e-beam evaporated a-Si films at high growth temperature suggested that increase in local mass density and structural order of the amorphous film can reduce the TLS and hence the cryogenic loss. Both reports pointed to that the change of local atomic arrangement of the amorphous state toward a more homogeneous and ordered configuration correlates with the reduction of the mechanical loss.

Techniques such as Extended X-ray Absorption Fine Structure (EXAFS) and Fluctuation Electron Microscopy (FEM) have recently been used to explore the atomic structure of tantala and Ti-doped tantala \cite{73_Bassiri2, 74_Bassiri3}. Theoretical modeling by using density functional theory molecular dynamic simulation with reverse Monte Carlo refinements ~\cite{75_Bassiri4, 76_Wu3} in conjunction with the experimental probing techniques will be useful to gain insight for the mechanical loss of the amorphous coating materials and will lead to feedbacks for material tailoring for improving coatings of the laser interference gravitational wave detector.

\subsection{Conclusion}

Ion beam sputter deposition is the method for depositing high quality amorphous coatings for laser interference gravitational wave detector. The equipment investment and operating cost is high for coating large-scale mirrors that hinders the rapid research request. Small-scale research type ion beam sputter deposition apparatus will be very useful for thin film material exploration. Chemical vapor deposition and the related materials are the alternatives that worth exploring due to its availability both for large and small-scale coatings with less hardware cost and matured technology know-how from the semiconductor IC industry. Ti-doped tantala and silica are the materials for the high and the low index thin films currently being used in the coatings of the room temperature laser interference gravitational wave detector. Doping level and heat treatment of these thin film materials are the major focus of coating research for improving the room temperature and the cryogenic temperature mechanical loss. Other high index materials such as hafnia, amorphous silicon, and silicon nitride showed promising aspects for improving coatings. Less effort for exploring new low index materials other than silica has been pursuit. New layer structures other than the conventional quarter-wave structure such as optimized layer structure, nano-layer structure, multi-material structure and Khalili cavity have shown some aspects of advantages over the conventional quarter-wave structure. Probing techniques such as Reduced Density Function electron diffraction, Extended X-ray Absorption Fine Structure, and Fluctuation Electron Microscopy have been applied to the Ti-doped tantala thin films and insights for the relationship between the mechanical loss and the atomic structure of the amorphous thin film materials have been gained. Various measurement techniques on the room and cryogenic mechanical loss \cite{23_Chao2,77_Nawodt,78_Cesarini,79_Nicol} and optical properties \cite{80_Vander, 81_Alexandrovski} have been developed for faster and reliable characterization of the thin film materials. Coating development for the laser interference gravitational wave detector is progressing in all the key areas.


\section{Advanced coatings for thermal-noise-free detectors}
\emph{Coating Brownian noise, driven by excess mechanical dissipation in high-reflectivity ion-beam sputtered (IBS) films imposes a severe limit on the performance of state-of-the-art precision measurement systems, such as cavity-stabilized lasers for optical atomic clocks and interferometric gravitational wave detectors.  Here we give an overview of the current thermal-noise-driven limitations in precision interferometry, introduce alternative solutions to minimizing these effects, and ultimately outline the path of development of our groundbreaking crystalline coating technology. We then describe this novel coating process in detail, show recent performance improvements and expanded application space, and finally discuss some remaining technical obstacles for employing our low-loss substrate-transferred crystalline coatings in next-generation gravitational wave interferometers.}

\subsection{Introduction}
Today's most advanced technologies for measuring time and space \cite{Schiller04}, particularly optical atomic clocks \cite{Ludlow15} and interferometric gravitational wave detectors \cite{Abbott09,11_Harry4}, are now encountering an ultimate barrier set by fundamental thermal processes. The current bounds of stability and sensitivity in these systems are dictated by the mechanical damping, and thus the corresponding Brownian noise level, of the high-reflectivity coatings that comprise the reflective elements of the cavity end mirrors. In these systems, thermally driven fluctuations result in modifications of the optical path length and hence unavoidable ``thermal noise'' \cite{Saulson90} in the interferometer. The need to minimize these effects has led to major advances in the design of optical cavities, which have continually redefined the state of the art in precision sensing \cite{Young99,Ludlow07,Millo09,Jiang11,Kessler12a,Nicholson12,Martin13}. Along these lines, interferometric gravitational wave detectors will soon be limited in their most sensitive measurement band by the thermal noise introduced by the lossy high-reflectivity dielectric multilayers \cite{LSC}. Thus, the most significant remaining impediment for achieving enhanced performance in these systems is the Brownian motion of the cavity's end mirrors \cite{Kessler12a,Numata04,Bishof13}. According to the fluctuation�dissipation theorem \cite{3_Callen,Landau96}, this motion is directly linked to mechanical damping in the constituent coating materials of the coating. Thus, the long-standing challenge has been to identify materials simultaneously capable of high reflectivity and low mechanical dissipation.

In order to overcome this limitation, a concerted effort has focused on the exploration of alternative reflector topologies and particularly on the identification of optical coatings capable of simultaneously achieving high-reflectivity and minimal mechanical dissipation. Building upon advances in quantum optomechanics \cite{Aspelmeyer14}, surface-emitting semiconductor lasers \cite{Iga00}, and advanced microfabrication techniques \cite{Madsen11}, we have made significant strides towards a novel solution, successfully integrating low-loss single-crystal multilayers with super-polished optical substrates. These ``crystalline coatings'' have demonstrated competitive optical properties with sputtered oxide films, while simultaneously exhibiting a tenfold reduction in mechanical loss angle at room temperature \cite{Cole13} and promise an additional order of magnitude reduction upon cooling to cryogenic temperatures \cite{Cole12}. Employing these novel epitaxial materials in a Fabry-Perot reference cavity, we have now realized excess optical losses (scatter plus absorption) as low as 3 ppm, with sub-ppm absorption measured at 1064 nm \cite{Cole15}. The exceptional performance of our coatings paves the way for the development of the next generation of long-baseline interferometers, unencumbered by coating thermal noise, and enabling the full exploitation of quantum noise reduction techniques in order to achieve unprecedented levels of strain sensitivity.

Dielectric multilayers based on metal-oxide thin films deposited by IBS have represented the state of the art in high-reflectivity optical coatings since the 1980s, being capable of optical scatter and absorption at the parts-per-million (ppm) level \cite{Rempe92}. The excellent optical quality of these reflectors has led to enormous progress in a broad range of applications including sub-attometer displacement sensitivities for gravitational wave observatories \cite{Abbott09} and frequency stabilities at the $10^{-16}$ level for metrology applications \cite{Jiang11,Kessler12a,Nicholson12}. In spite of their superior optical properties, the amorphous thin films at the heart of these coatings exhibit excess mechanical damping \cite{Crooks02,4_Harry1} driven by the internal losses in the high-index tantala (Ta$_2$O$_5$) layers \cite{40_Penn3}. Even with more than a decade of investigation, the mechanical damping of existing dielectric multilayer mirrors has remained unacceptably large, with a maximum reduction in the overall loss by a factor of two, with the best reported value to date for IBS-deposited dielectric multilayers of $2\times10^{-4}$ through the incorporation of a small fraction of TiO$_2$ into the high-index tantala layers \cite{52_Harry7}. This relatively large loss angle results in significant displacement fluctuations of the mirror surface arising from thermally driven mechanical modes, typically referred to as `coating thermal noise'\cite{11_Harry4} having been experimentally identified in the study of gravitational wave detectors \cite{Crooks02,4_Harry1,40_Penn3}. Over the course of the preceding decade only modest improvements in the loss angle have been realized, leading to the investigation of a number of alternative solutions.

\subsection{Alternative solutions}
Following the initial experimental verification of excess damping by the LIGO scientific collaboration, the minimization of coating thermal noise has been a long-standing challenge in the precision measurement community. As described above, until recently, progress has been modest. Even with the implementation of improved coatings based on titania-alloyed tantala, the Brownian noise of the high reflectivity multilayer remains the dominant noise source for optical reference cavities, and, coupled with quantum noise effects, for interferometric gravitational wave detectors as well. In lieu of further improvements to the coating thermal noise performance, many researchers have investigated modified interferometer designs that minimize the overall sensitivity to thermal noise \cite{Amairi13}. This can be achieved for example by increasing the size of the optical mode \cite{Amairi13,Bondarescu08}, as sampling a larger area of the mirror surface effectively averages out small-scale fluctuations, or by exploiting the coherent character of the underlying displacements and strains for potential cancellation \cite{Kimble08}. In a similar vein, various proposals have been put forward to significantly alter or even eliminate the coating entirely, including resonant waveguide grating reflectors \cite{Friedrich11}, photonic crystal reflectors \cite{Kemiktarak12}, or via total-internal-reflection-based cavities. These approaches, however, have not yet demonstrated either a sufficiently high mechanical quality and sufficiently narrow cavity linewidth (either by unavoidable thermorefractive noise in the case of whispering gallery mode resonators \cite{Alnis11}) or a sufficiently low thermal-noise performance. The most effective improvements in the interferometer thermal noise performance have resulted from the implementation of components (i.e. substrates and suspensions) constructed from materials with minimal mechanical damping. Ultimately, further performance enhancements in the next generation of precision interferometers will only be possible through solutions that address the limiting mechanical loss of the coating itself. Moreover, as with recent developments in optical reference cavities \cite{Kessler12a}, future systems will likely seek to minimize thermal noise effects through a reduction in the operating temperature, requiring excellent optical and mechanical performance at low temperature as well.

\subsection{Crystalline coatings}
Recent work in quantum optomechanics has provided a com-

\noindent pletely novel solution for the reduction of coating thermal
noise. This field of research exploits optomechanical interactions within optical cavities to control and study the quantum regime of nano- to macro-scale mechanical oscillators \cite{Aspelmeyer14}. Similar to the requirements found in an ultrastable interferometer, reaching the quantum regime of mechanical motion necessitates the implementation of structures with both high optical and mechanical quality. In this context, monocrystalline Al$_x$Ga$_{1-x}$As heterostructures (AlGaAs) have been identified as a promising option for multilayer mirrors, in particular, as these structures are capable of significantly reduced loss angles while maintaining excellent optical performance \cite{Cole12,Cole08,Cole10b}. Behind silicon, AlGaAs is the most explored semiconductor material and is thus the most mature option for producing high-quality single-crystal Bragg stacks, consisting in this case of lattice-matched ternary alloys of GaAs and AlAs. These materials may be epitaxially grown as single-crystal heterostructures on a suitable lattice matched substrate, enabling the production of arbitrary stacks of high index-contrast materials that maintain nearly perfect crystalline order. While AlGaAs-based distributed Bragg reflectors (DBRs) have been applied for the fabrication of optical interference coatings since the mid 1970s \cite{Vanderziel75}, until 2007, the mechanical damping of this materials system had not been methodically investigated.

Recently, measurements of free-standing mechanical resonators microfabricated directly from epitaxial AlGaAs multilayers yielded exceptional quality factors, $Q$, up to 40000 at room temperature \cite{Cole12}. For an isolated resonance, $Q$ can be converted to loss angle via $\phi = Q^{-1}$, demonstrating loss angles as low as $\sim2.5\times10^{-5}$ at room temperature. In comparison, similar measurements of free-standing SiO$_2$/Ta$_2$O$_5$, the typical choice for high reflectivity mirror structures, fall in the range of a few thousand \cite{Brodoceanu10}, yielding values consistent with the coating loss angles observed in studies of optical reference cavities and gravitational wave detectors \cite{Crooks02,4_Harry1,40_Penn3}. Taking into account the competitive optical performance of AlGaAs-based DBRs, together with the potential for ppm-levels of scatter in absorption in the near infrared, these epitaxial multilayers represent a promising alternative for the development of mirrors with ultralow thermal noise. After our initial demonstration of low-loss micro-resonators, the most pressing question was whether such structures could be successfully scaled up and applied as ``macroscopic'' cavity end mirrors.

\subsection{Fabrication process and performance}

Preliminary research in the field of cavity optomechanics revealed that AlGaAs-based heterostructures are capable of significantly reduced mechanical damping, while achieving competitive reflectivity, when compared with state-of-the art ion-beam sputtered dielectric coatings \cite{Cole12,Cole08,Cole10b,Cole10a,Cole11}. Although excellent optomechanical properties had been demonstrated in suspended micrometer-scale resonators, the implementation of such single-crystal multilayers in a high-finesse cavity presents a number of challenges. Most importantly, the choice of substrate materials is quite limited with these coatings. Direct deposition onto typical optical substrates is precluded by lattice matching constraints, or in the case of amorphous substrates, by the lack of a crystalline template for seeded growth. An additional difficulty arises as stable optical cavities require curved mirrors, the realization of which is incompatible with the current capabilities of high-quality epitaxial film growth.

Such limitations are not found with ion beam sputtering. With this process, high quality multilayers can be deposited on essentially any relevant, and even structured, substrate assuming sufficient surface quality and adhesion can be realized. Nonetheless, by exploiting advanced semiconductor microfabrication processes, we have overcome these obstacles and realized the successful implementation of this low-loss materials system in a standard optical cavity configuration. Rather than developing complex crystal growth processes to realize the direct deposition of AlGaAs multilayers, we have instead developed a technique for the transfer of low-loss monocrystalline multilayers onto essentially arbitrary optical surfaces.

Ultimately, this substrate-transfer coating process entails separating the epitaxial multilayer from the original growth substrate and directly bonding it---using no adhesives or intermediate films---to the desired (curved or planar) optical\linebreak
\vspace*{-4mm}

\begin{figure}[H]
  \centering
  \includegraphics[scale=1]{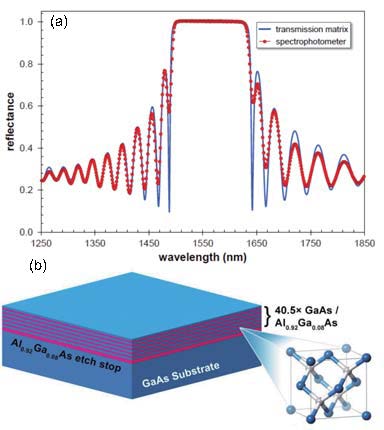}
  \caption{(Color online) (a) Fitted reflectance spectrum (red points, spectrophotometer measurements; blue line, transmission matrix theory) of an example 1550 nm GaAs/AlGaAs single-crystal multilayer. An excellent fit is achieved, with the absolute reflectance value limited by the wavelength resolution (1 nm) of the instrument. (b) Cross-sectional schematic of a representative crystalline multilayer. Inset: the zincblende unit cell. The mirror consists of alternating quarter-wave GaAs (high index, 3.38 at 1550 nm) and Al$_{0.92}$Ga$_{0.08}$As (low index, 2.93 at 1550 nm) grown on a (100)-oriented GaAs substrate via MBE. The base of the mirror incorporates a thick Al$_{0.92}$Ga$_{0.08}$As etch stop layer for protection during the substrate removal process.}\label{Fig1}
\end{figure}
\begin{figure}[H]
\centering
\includegraphics[scale=1]{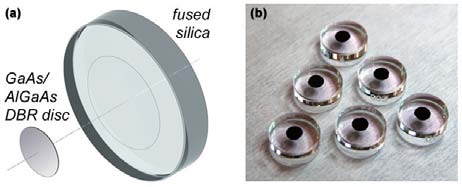}
\caption{(Color online) (a) Exploded view of a bonded mirror assembly showing the GaAs/AlGaAs mirror disc and fused-silica substrate with a polish-imparted 1 m radius of curvature (ROC). (b) Photograph of a set of completed 16-mm diameter mirrors. These devices incorporate a 5-mm-diameter crystalline coating transferred to a 6-mm thick fused-silica substrate with a 1 m ROC.}
\label{fig2}
\end{figure}


\noindent substrate. In terms of the microfabrication details, this process can be seen as an extension of foundational work such on semiconductor ``wafer fusion'' \cite{Black97}, as well as early demonstrations of epitaxial layer transfer \cite{Konagai78,Yablonovitch89}, and more recent stamp-mediated methods \cite{Madsen11} and, at a very basic level, is analogous to optical contacting, a widely used approach for the construction of optical subassemblies. The most significant difference here is that chemical treatment of the bonded interface, coupled with an appropriate annealing process can yield interfacial strengths on par with inter-atomic/bulk bond energies. The fabrication process begins with the deposition of a high-quality and high reflectivity (Figure~\ref{Fig1}(a)) epitaxial multilayer on a GaAs substrate (Figure~\ref{fig2}(b)). In this case the multilayer is grown using molecular beam epitaxy (MBE), enabling the highest optical quality material, particularly with respect to background impurities and thus optical absorption. The multilayer itself consists alternating quarter-wave optical thickness layers of high and low-index materials generated by modulating the Al content of the constituent films. This simple Bragg structure exhibits a low-transmission stopband centered at the desired operating wavelength, typically 1064 or 1550 nm. With the multilayer completed, definition of the AlGaAs mirror disc relies on optical lithography, in order to define the lateral geometry of the disc, followed by chemical etching to extrude the disc shape through the deposited films and into the substrate. Chemo-mechanical substrate removal is then used to strip the GaAs growth template. Next, the thick AlGaAs etch stop layer is removed and the mirror surface is cleaned of any potential debris. Finally, the crystalline mirror disc and the silica substrate are pressed into contact, resulting in a spontaneous van der Waals bond. To strengthen the interface and minimize potential frictional losses, a post-bond anneal completes the fabrication procedure.

\begin{figure}[H]
  \centering
  \includegraphics[scale=1]{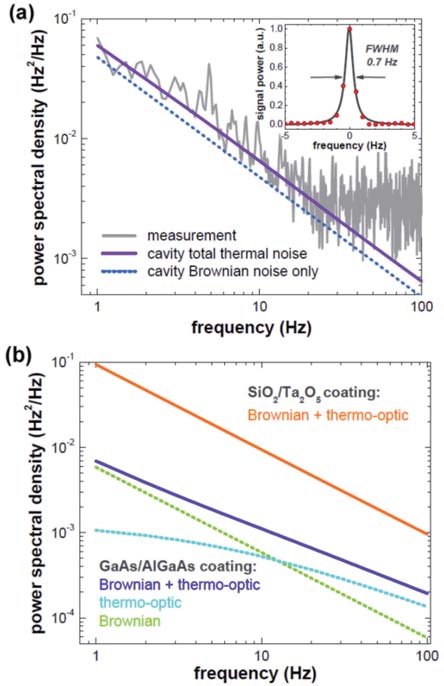}
  \caption{(Color online) Noise performance of a 35-mm long reference cavity utilizing crystalline coatings. (a) Thermal-noise-limited noise power spectral density (NPSD) of a cavity-stabilized 1064-nm laser system, with a 0.7 Hz linewidth (0.5 Hz resolution bandwidth) shown in the inset. (b) Comparison of the coating noise performance for our crystalline multilayer and a typical dielectric coating. At 1 Hz the GaAs/AlGaAs crystalline coating yields more than a tenfold improvement in Brownian noise.}\label{Fig3}
\end{figure}

Using this technique, we circumvent the impediments arising from the direct deposition of monocrystalline multilayers onto arbitrary surfaces (Figure~\ref{fig2}(a)) and realize high-quality compound semiconductor multilayers transferred to planar and curved super-polished substrates. A photograph of completed mirrors is presented in Figure~\ref{fig2}(b). With this process we can implement epitaxial semiconductor materials as high-quality optical coatings for the first time, enabling the construction of optical cavities capable of significantly improved thermal noise performance when compared with similar structures employing sputtered dielectric mirrors. In a seminal experiment published in 2013 \cite{Cole13}, our crystalline coatings were employed as end mirrors in a compact optical reference cavity, exhibiting ppm-levels of optical losses (with a demonstrated finesse of 150000 at 1064 nm). Even more importantly, a thermally-limited noise floor consistent with a tenfold reduction in mechanical damping, with an upper limit on the loss angle of $4\times10^{-5}$, was observed at room temperature \cite{Cole13}. This represents an order of magnitude reduction in the mechanical dissipation when compared with typical ion-beams sputtered films, as shown in Figure~\ref{Fig3}, and based on measurements of microfabricated resonators at 10~K

\noindent promises the possibility for a further order of magnitude improvement at cryogenic temperatures \cite{Cole12}. Thus, coating Brownian noise levels can ultimately be reduced by up to a factor of ten with our mirrors, leading to significant performance enhancements in precision optical interferometers requiring the utmost stability.

Beyond Brownian noise, an independent source of significant coating-related noise arises from thermal fluctuations in the multilayer and substrate driven by the finite thermal expansion coefficient \cite{9Braginsky2}, as well as through the temperature dependence of the index of refraction of the constituent films \cite{9Braginsky2}, referred to as thermo-elastic and thermo-refractive noise, respectively or together as thermo-optic (TO) noise. The majority of the epitaxial multilayers thus far employed in our crystalline coating experiments are standard quarter-wave structures and given the large thermal expansion and thermorefractive coefficients of AlGaAs results in significant TO-noise at high frequencies. Fortunately, in stark contrast to Brownian noise, the components of TO noise can be made to add coherently and thus can in principle be eliminated by careful design of the layer structure of the mirror \cite{Evans08,6_Gorodetsky}. As a first step in this direction, we have recently demonstrated a custom-tailored multilayer using an advanced optimization procedure capable of successful coherent cancellation of TO noise in a crystalline GaAs/AlGaAs multilayer \cite{Chalermsongsak15}.

In summary, our direct bonded `crystalline coatings' represent an entirely new paradigm in optical coating technology, exhibiting both intrinsically low mechanical dissipation (and thus reduced Brownian noise), as well as high optical quality, on par with sputtered oxide reflectors. It is important to realize that these characteristics can be realized in a standard configuration without the need for significant changes to the overall interferometer design, unlike the use of gratings or total internal reflection techniques.

\subsection{Improved optical performance}

Following the initial demonstration described above, a significant effort has been undertaken to improve the optical performance of our low-loss crystalline coatings through optimization of the crystal growth and substrate-transfer processes. Previously, AlGaAs multilayers have exhibited typical excess losses (scatter + absorption) at the $\sim20$ ppm level \cite{Cole13}. With an in-depth focus on minimizing the background impurity level of the constituent films over the last few years, we can now achieve an optical absorption level at or below 1 ppm in the near infrared, between 1000--1600 nm. Most recently, by improving the quality of the substrate-transfer process, we have minimized optical scatter losses, reaching limiting levels of $\sim3$ ppm in the same wavelength range. Putting these improvements together, we have now realized crystalline-coated cavity end mirrors bonded to super-polished fused silica substrates capable of excess loss levels below 5 ppm \cite{Cole15}. With a transmission of 10 ppm at a center wavelength of 1550 nm, these coatings are capable of finesse values exceeding $2\times10^5$, while a reduction in transmission to 5 ppm enables a finesse of $3\times10^5$. These results represent a significant enhancement in optical quality and
now prove that crystalline coatings are capable of achieving optical loss values on par with those found in high-quality IBS coatings.

Follow-on investigations have now shown the potential for realizing parts-per-million levels of optical losses, including both absorption and scatter, in GaAs-based Bragg mirrors at wavelengths spanning 1000 to nearly 4000 nm. In collaboration with colleagues from the LIGO-scientific collaboration, we have experimentally verified absorption coefficients below 0.1 cm$^{-1}$ in the near infrared \cite{Steinlechner15}.
These recent advancements have opened up additional application areas including but not limited to crystalline coatings for next-generation ring-laser gyroscopes \cite{Schreiber15}, as well as chemical and trace gas sensing in the mid-infrared spectral range. Further efforts include a focus on increasing the current maximum bond diameter of 16 mm \cite{Steinlechner15}, aiming for tens-of-cm-diameter GW-relevant optics.

\subsection{Summary and path forward}
With the introduction of low-loss and high-reflectivity end mirrors based on substrate-transferred crystalline coatings, we have demonstrated unprecedentedly low Brownian noise in high-reflectivity optical coatings \cite{Cole13}. The observed tenfold reduction in coating loss angle, compared with state-of-the-art IBS-deposited thin films, represents a long-awaited breakthrough for the precision measurement community. In the field of cavity-stabilized laser systems, combining our crystalline coatings with optimized cavity designs will result in an immediate enhancement of the achievable frequency stability of narrow-linewidth lasers, opening up the $10^{-17}$ stability regime at room temperature. Furthermore, the demonstrated reduction in thermal noise provides a path towards ultrastable, compact and portable laser systems and optical atomic clocks. Going forward, we anticipate further improvements in the optomechanical performance of these mirrors. As opposed to dielectric multilayers where the $Q$ saturates or even decreases at low temperatures \cite{Yamamoto06}, measurements of free-standing epitaxial multilayers in a cryogenic environment reveal loss angles nearly another order of magnitude lower, down to $4.5 \times10^{-6}$ ($Q$ of $2.2\times10^5$) at 10 K \cite{Cole12}, promising substantial improvements in the thermal noise performance of next-generation cryogenic interferometers.

Looking forward towards the development of relevant mirror sizes for interferometric gravitational wave detectors (with coating diameters $>20$ cm), we find that the current commercially available GaAs substrate size is the most serious limitation. Recently, we have successfully demonstrated high-yield bonding of GaAs to fused silica at a diameter of 10 cm, with the process being immediately transferrable to sizes up to 20 cm. Moving beyond the maximum commercially available substrate diameters of 20 cm would require the development of single-crystal GaAs boules of increased diameters, even beyond 30 cm. The other potential path forward would be to move to heteroeptixial growth of the multilayer. In this vein, it may also be possible to realize high quality AlGaAs structures on germanium wafers or on engineered SiGe alloy structures grown initially on a single-crystal silicon substrate \cite{Ting00}, although this hetero-epitaxial system is still admittedly in the nascent stages of research and exhibits significantly diminished material quality when compared with homo-epitaxial growth of AlGaAs on GaAs. Both of these directions represent a significant engineering challenge, however in principle, with our microfabrication-based substrate transfer process, we foresee no fundamental barriers to realizing mirror sizes relevant for interferometric gravitational wave detectors.


\section{Optical properties of test masses for next generation gravitational wave detectors}
\emph{The main optical properties of the test masses for the third generation of the gravitational wave detectors are presented. For room temperature interferometer, well known fused silica is still considered the best substrate. For low temperature interferometers, sapphire and silicon, crystalline materials, available in large ultra-pure crystals are the most promising candidates.
}

\subsection{Introduction}\label{sec:intro}

The second generation laser gravitational wave detectors such as Advanced LIGO \cite{LIGO_intro}, Advanced Virgo \cite{Virgo_intro} or KAGRA \cite{Kagra_intro} are giant sophisticated Michelson interferometers. The two perpendicular arms of the Michelson are replaced with kilometer long Fabry-Perot cavities to increase the interaction time between the interferometer circulating light and the gravitational wave signal. Due to their essential roles, the arm cavities are the most critical optical system of the detectors and the large mirrors forming the cavities are commonly called test masses.

In this section, the most important optical characteristic of the test masses are described. The properties of fused silica are detailed for room temperature interferometer and the research toward sapphire and silicon large test masses is presented.


\subsection{Requirement for test masses}\vspace*{-1mm}

Before reviewing the different test masses materials in the next sections, we will first detail the essential properties to consider while choosing the test mass materials for gravitational wave detectors.

\subsubsection{The optical absorption}

The optical absorption of the test mass substrate is one of the most critical parameters because some substrates such as the input mirrors of the arm cavities or the beamsplitter are used in transmission and so are crossed with the high power laser beam. As a consequence of the optical absorption, a small fraction of the laser light will be absorbed and then converted into heat. The consequences of the heat generation are different if the test masses are at room temperature or at low temperature:

  (1) At room temperature, the heat generated by the absorbed laser beam will create a gradient of temperature, inducing then a gradient of refractive index by the thermorefractive effect (the change of the refractive index with the temperature). The wavefront distortion from the gradient of refractive index, is usually equivalent to a convergent lens and hence this effect is often called thermal lensing \cite{TL1}.
      Thermal lensing can seriously limit the performance of the detectors by degrading the control signals and reducing the coupling between cavities (called mode matching). For the second generation interferometers, an extensive and complex thermal compensation system has to be implemented in order for the interferometers to reach their sensitivities \cite{TL2}.

  (2) At low temperature, the problem is different, due to the high thermal conductivity \cite{TL3} and low thermorefractive coefficient \cite{TL4} of the substrate, the thermal gradient is greatly reduced, and the hence the induced wavefront distortions is negligible. The problem in that case is due to the rise of temperature due to the heat absorbed. Indeed, it is very difficult for the heat to escape due to the inefficiency of the radiative cooling at low temperature and since the mirror is only suspended by very thin long fiber, poorly conducting the heat away \cite{TL5}.
      So if the optical absorption is not low enough, it may be impossible to cool down the test mass to their operating temperature.

It is worth noting that the optical absorption originated both from the substrate and from the coating. For room temperature second generation interferometers, the coating is the main contributor of thermal lensing due to the very high optical power circulating in the arm cavities (close to a MegaWatt of light and seen only by the high reflectivity coating of the test masses).

Even if the power absorbed could have dramatic consequences, one must be aware that the optical absorption is a negligible direct source of optical loss in the interferometer.

\subsubsection{The mechanical loss}

Another essential property to consider is the mechanical loss that is directly related to the level of the Brownian displacement noise. That is not an optical property, so it will be not detailed here but a dedicated article is present in sect. 1.

\subsubsection{The availability in large size}

In order to reduce the magnitude of the thermal lensing and the level of the thermal noise, it is recommended to have large beams (several centimeters of radius) on the mirrors. Moreover, because of the diffraction of the laser beam as the arm cavities goes longer, the size of the laser beam inevitably increases on the input and end mirrors of the arm cavities. As a consequence of large beams, large mirrors are required with at least a diameter three times the diameter of the laser beam \cite{CL_loss}.

Not only the diameter is important but also the total weight of the optic has to be taken into account. Heavy optics are advantageous to reduce the radiation pressure noise \cite{RP}. As an example, for Advanced LIGO and Advanced Virgo, the test mass are made of fused silica, with a diameter of 350~mm, thickness of 200~mm and for a total weight of 40~kg.

\subsubsection{Other considerations}

Once the test mass substrate is produced by the glass manufacturer, two steps still remain before having one arm cavity mirror which could be then installed in the interferometer. First, the substrate has to be polished to have the right profile (with ideally a perfect spherical shape) and second, coatings have to be deposited, usually one high reflectivity coating on one side and one anti-reflection coating on the other side.

So the test mass material for the next generation interferometers has also to be compatible with the last two steps. For example, sapphire, a potential substrate material for low temperature interferometer has been known to be difficult to polish due to its hardness, leading to extra-loss in the arm cavities (exta light scattered). This problem is now solved thanks to new techniques of polishing \cite{22Hirose}.

\subsubsection{Summary}

As we saw in the previous part, the choice of the substrate material for the test masses of the gravitational wave detectors is seriously constrained. Hence, only very few materials are suitable to be used for the large optics of the detectors. In the next sections, we will review the three most common ones.

\vspace*{-1mm}
\subsection{The supremacy of fused silica at room temperature}\vspace*{-1mm}

Fused silica is a well known material since it was already the test mass material for the first and second generations of gravitational wave detectors. It is a well justified choice since the fused silica used has some outstanding optical and mechanical properties.

The test masses of Advanced LIGO and Advanced Virgo were made of Suprasil 3001 from the manufacturer Heraeus. The material is a high purity synthetic fused silica materials manufactured by flame hydrolysis of SiCl$_4$. The material is free from bubble and inclusions and has a extremely high uniformity. As an example, on the input mirrors of Advanced Virgo, the relative change of the refractive index over the central 200 mm diameter is less than 1 part per million (ppm). The birefringence is also well under control with value around 1 nm/cm on the same central diameter.

The fused silica used as a test mass is a special grade of fused silica with a reduced content of OH, as the result the optical absorption measured at 1064~nm is also impressively low. To give a number, the volume absorption of the fused silica test masses is routinely measured below 0.5~ppm/cm$^{1)}$.

A quick note on thermal noise, the fused silica bulk presents very low mechanical loss at room temperature \cite{19Penn} and hence has a very low thermal noise. Furthermore, fused silica can be molded/machined with arbitrary shape and then the attachment system of the test mass to the suspension can also be made of fused silica. This monolithic suspension as it is called can reduce the thermal noise suspension \cite{FS_susen}.

\begin{figure}[H]
\centering
\includegraphics{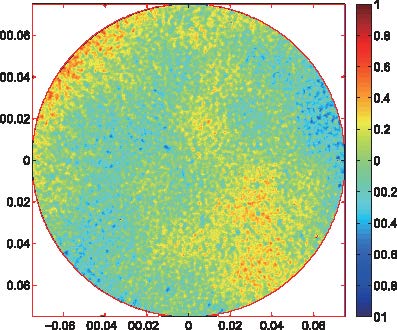}
\caption{(Color online) Example of one the polished test mass surface of Advanced Virgo. The color scale represents the height of the surface in nanometer over the central diameter of 150~mm after the curvature has been substracted. The polishing is really outstanding with a surface height RMS of 0.17~nm.}
\label{fig:figure1}
\end{figure}

Thanks to the long experience working with fused silica, the polishing is also not an issue. For the second generation interferometers, extremely good polishing has been achieved. For example, for Advanced Virgo, the low spatial frequency figure (also called flatness) was below 0.3~nm RMS over the central part as shown in Figure \ref{fig:figure1}. The high spatial frequency of the defects (called microroughness) was below 0.1~nm RMS. From the optical point of view, after polishing the test masses could almost be considered as perfect. It is likely that those numbers will be similar for the third generation of interferometer.

What makes this fused silica even more remarkable is that it is available in large size. Diameter of more than half a meter are possible, as demonstrated in Advanced Virgo where the beam splitter has a diameter of 550~mm and weights 40~kg.

As shown in this section, fused silica is an amazing material with no competitor as a test mass material. As such, it is highly likely that fused silica will still be used for the next generation of room temperature interferometers. However, it should be noted that fused silica becomes a lossy mechanical

\end{multicols}
\vspace*{-2mm}
\noindent\rule{2.5cm}{0.4pt}\\[0.1mm]{\qihao \hspace*{4mm}1) Personal communication, Forest D, LMA, 2015}

\begin{multicols}{2}
\renewcommand{\baselinestretch}{1.08} \baselineskip 12.2pt\parindent=10.8pt

\noindent material as the temperature reduces and hence is not considered for low temperature interferometers.

\subsection{The coming back of sapphire}

The sapphire is not a new comer as test mass substrate, at the turn of the century sapphire was a serious candidate for the test masses for Advanced LIGO \cite{A_LIGO_sap}. Finally, fused silica won the match but the game was tight. Compared to glass, sapphire has a very high thermal conductivity (46~W/mK, more than 30 times the ones of fused silica), that is very advantageous to reduce the magnitude of the thermal lensing, however it comes with a price: a high thermoelastic noise at room temperature. Therefore nowadays sapphire is only considered for low temperature gravitational wave detectors.

Sapphire is a crystalline material suitable for low temperature interferometer. Sapphire will be used by the Japanese detector KAGRA with the sapphire test masses cooled down to 20~K. As such much of the current research done on sapphire for large mirrors is done within the framework of KAGRA.

One drawback of sapphire is its relatively high absorption. Absorption of 50~ppm/cm at 1064~nm is common. The absorption has both been measured at room temperature \cite{sap_abs} and cryogenic temperature \cite{sap_abs_cryo} and presents the same magnitude. Such a level of absorption has some serious consequences at low temperature since it limits the amount of power circulating in the interferometer and also puts constraints on the suspension design to allow the evacuation of the excess heat. For example in KAGRA, it is expected that 1~W will be absorbed by the test masses of the input cavities, this power may seem small but it requires already larger than desired diameter suspension wire. Similarly to what has been achieved with fused silica, the lower stage of the KAGRA suspension will all be made in sapphire \cite{sap_sus_cryo}.

Currently sapphire test masses is available up to a diameter of 220~mm and recently, polishing as good as what is achieved with fused silica has been demonstrated \cite{22Hirose}.

\subsection{The large outsider: silicon}

For the third generation detectors, it is planned to have even longer arm cavities ranging from 10~km for the Einstein Telescope (ET) \cite{ET_bib} to 40~km for a potential long term successor of Advanced LIGO. Due to the divergence of the laser beam, longer arm cavities mean bigger beam spots on the test masses and so it requires larger mirrors.
Fused silica is still the favorite material candidate for such room temperature cavities since test masses with diameter above 500~mm are possible. However at low temperature, sapphire is not available on such a large diameter, hence an alternative material has to be found.

\begin{figure}[H]
\centering
\includegraphics{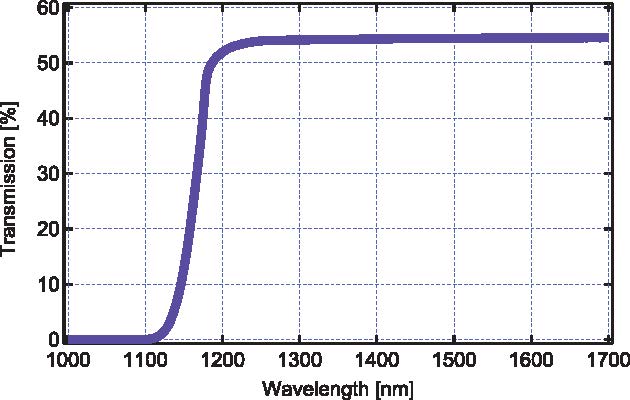}
\caption{(Color online) Transmission spectra of a silicon window. Silicon is a semiconductor with a band gap of 1.1 eV and hence the transparency region only starts above 1200~nm. The raw transmission of the sample is only 55\%, not because of the absorption but because of the Fresnel reflection on each side of the window. Silicon has a very high index of refraction: 3.45, and so reflection per uncoated side is around 30\% (compared to 4\% for fused silica).}
\label{fig:figure2}
\end{figure}

This new material is mono-crystalline silicon, the material king of the semiconductor industry. It is available in very high purity (i.e. without doping) and also in large size as the industry is moving toward 450~mm diameter wafers. As shown in Figure \ref{fig:figure2}, silicon is not transparent at 1064~nm at the opposite of fused silica and sapphire, so a new wavelength has to be used for the interferometer, the choice went for 1550~nm, since laser and optical components already exist at this wavelength.

The optical absorption of silicon at 1550~nm has been found to be directly related to the concentration of impurities implemented in order to lower the resistivity of the material. For high purity silicon, with impurities concentration below 10$^{12}$ cm$^{-3}$, the absorption has been measured below 10~ppm/cm \cite{abs_si1}. Such low absorption silicon is produced by the so-called 'float zone' growth technique and is unfortunately only available for diameter below 200~mm. Larger diameter silicon is produced with the Czochralski process technique, leading to samples with higher impurities concentration and hence higher optical absorption (around 1000~ppm/cm at best).

At room temperature, the near infrared optical absorption of silicon is due to the presence of free carrier in the conduction or valence bands. Every doping atom is ionized and so creates one free carrier, hence the concentration of free carrier is equal to the concentration of doping atom. At low temperature, when there is not enough thermal energy to ionized the dopant, the concentration of free carrier is nil. So it was believed that the silicon absorption could be very low at cryogenic temperature. Measurement have shown that it is not the case and that absorption at low temperature has the same magnitude as at room temperature. Indeed, it is understood that at low temperature, the absorption is due to ionisation of the neutral dopant (that is the near infrared tail of photoionization process).

So as a conclusion regarding the absorption of silicon at low temperature, it is likely that for large diameter optics made with the Czochralski process the absorption will be \linebreak
\vspace*{-2mm}

\noindent relatively high (in the order of 1000 ppm/cm). It is is not a showstopper, but it puts serious constraints on the circulating power in the interferometer and on the cooling power of the cryogenic suspension.

The polishing of silicon is not an issue since surface flatness as good as for fused silica can be achieved. Reports from the polishers mentioned that the roughness (the high spatial frequency) is higher compared to fused silica and silicon is relatively brittle, so the optics must be manipulated with care.

\vspace*{-1mm}
\subsection{Conclusion}\vspace*{-1mm}

As explained in this section, fused silica will still keep its supremacy as the material of choice for the next generation of room temperature gravitational wave detectors. It is a well mastered material with outstanding optical properties and, as important, also available in large size.

For low temperature interferometers, two mono-crystalline materials are foreseen for the main mirrors: sapphire and silicon. Sapphire will be demonstrated in KAGRA in the coming years and for the long term silicon is planed. However, for silicon, large substrates with low optical absorption still have to be demonstrated.

In this section, optical properties of test mass material have been discussed. One should not forget that as important is the coating optical properties as coating could increased the overall absorbed optical power and be also the dominant source of  thermal noise as the second generation room temperature gravitational wave detectors.


\section{Lasers for next generation gravitational wave detectors}
\subsection{Introduction}\label{introduction}

Interferometric gravitational wave detectors (GWDs) sense differential arm length variations by the evaluation of the interference of two light beams that traveled along perpendicular interferometer arms. Both beams originate from the same single-frequency laser light source, which has to fulfill several highly demanding requirements. Single-frequency laser sources with several hundred Watt of output power are typically required that emit a linear polarized beam in a pure spatial mode. Furthermore the laser light has to be extremely stable.  Even though high attention is paid by the laser manufacturer to a low noise design, currently available laser show orders of magnitude too high fluctuations in several of their parameters to be used in next generation GWDs. Hence the light has to be actively and passively stabilized before injection into the interferometer.
Conceptually this stabilization is often split into a pre-stabilization performed on an optical table in air and into an in-vacuum part, which utilized the passive filtering by a suspended several tens of meter long optical cavity called input-mode-cleaner (IMC).
This section will review the laser requirements and possible candidates for the pre-stabilized laser (PSL) for next generation GWDs.

\subsection{Laser light sources for next generation interferometric gravitational wave detectors}\label{laser}

Currently several design path for next generation GWDs are followed, which require various types of lasers. The European design study for ET \cite{ET2009} for example is based on two different laser sources. The interferometer optimized for high GW signal frequencies needs a $500\,{\rm W}$ light source at a wavelength of $1064\,{\rm nm}$. To reduce the coupling of thermal noise into the GW readout channel, the spatial beam profile of this laser should resemble a Laguerre Gauss $LG_{33}$ mode. The second ET interferometer that is optimized for low signal frequencies is based on a  $3\,{\rm W}$ laser beam in a fundamental Gaussiam spatial mode at $1550\,{\rm nm}$. Other design efforts consider next generation GWD that operate at different wavelength and with different spatial profiles \cite{Smith2014, Freise2011}.
Hence the GWD laser research currently concentrates on investigations to test the feasibility of the proposed light sources. Several design consideration have to be taken into account. The most important ones are related to the wavelength, outpower and beam shape. Furthermore the spatial purity of the laser has to be high to allow for an efficient use of the light power in the optical cavities of the GWD. Hence a sophisticated thermal management of the laser is required as effects like thermal lensing or bi-focusing can easily degrade the spatial beam profile. Similarly depolarization effects have to be avoided or compensated to allow for the generation of a high power beam in a linear polarization state. Single frequency operation as well as a low noise design is required and design choices should allow for a laser design with high reliability and low costs.

One option to generate a light source for future generation GWDs is to modify second generation GWD light sources that have already proven their high reliability. Two experiments following this approach started with the high power beam of an Advanced LIGO type PSL \cite{Kwee2012}.
In a first experiment \cite{Meier2010} the output beam of the Advanced LIGO PSL was injected into an optical cavity resonant at $1064\,{\rm nm}$ that included a lithium triborate (LBO) crystal. A $149\,{\rm W}$ beam injected into this resonator was converted via second-harmonic generation into a laser beam at $532\,{\rm nm}$ with $134\,{\rm W}$ and a higher order mode (HOM) content of less than 3\%.
In a second experiment \cite{Carbone2013} the light from the Advanced LIGO PSL with a power of $140\,{\rm W}$ in the fundamental Gaussian mode was aligned onto a diffractive phase plate and converted into a Gauss Laguerre $LG_{33}$ mode. After spatial filtering of this beam by a resonant linear optical cavity a pure $LG_{33}$ mode with a power of $83\,{\rm W}$ was generated. The mode purity was estimated to be above 95\% and did not show any significant power dependence.

Another promising design of a laser source for next generation GWDs is based on the fiber technology. To reduce non-linear effects like stimulated Brillouin scattering fiber designs with  mode field diameter of approximately $20\,{\rm \upmu m}$ are used. Intentionally introduced losses for higher order spatial fiber modes can be utilized to achieve a beam profile that is close to the fundamental Gaussian mode.
The fiber amplifier system described in ref. \cite{Theeg2012} with a power of $300\,{\rm W}$ at a wavelength of $1064\,{\rm nm}$  can be used as an example to discusse the advantages and challenges of high power fiber amplifiers. As in most currently operating GWD lasers an non-planar ring-oscillator (NPRO, \cite {Kane1985a}) with a power of $2\,{\rm W}$ is used to seed a fiber pre-amplifier, which generates $25\,{\rm W}$ output power. This beam is then ampified by a high power fiber amplifier to an output power of $300\,{\rm W}$. The high power stage is an all-fiber design in which the pumplight provided by four fiber coupled laser diodes is coupled into the active large mode area (LMA) fiber via a $4+1 \times 1$ combiner \cite{Theeg2012a}. Hence no alignment drift is to be expected in the high power stage. The fiber design furthermore allows for a very high optical efficiency and hence low fabrication and operational costs. Care has to be taken in high power fiber amplifiers to avoid non-linear effects, in particular stimulated Brillouin scattering (SBS).
Furthermore the high intensities at the fiber/air interface and pump light coupled from the fiber cladding into the fiber coating constitute risks for fiber damage. Currently longterm stable operation of single-frequency linear-polarized high power fiber amplifiers is still to be demonstrated.

Only little experience exists in the GW community with the operation and stabilization of high power laser systems at $1550\,{\rm nm}$.
Laser with the output power required for the low power ET interferometer are commercially available and Er$^{3+}$ based fiber amplifiers are promising candidates for higher power levels. Pure Er$^{3+}$ fibers as well as Yb co-doped Er$^{3+}:{\rm Yb}^{3+}$ fibers have been operated at output power levels beyond $50\,{\rm{W}}$ \cite{Steinke2014, Kuhn2011}. Further research is required to increase these power levels and to optimize the spatial beam profile of the pure Er$^{3+}$ fiber amplifier presented in ref. \cite{Kuhn2011}. The Er$^{3+}:{\rm Yb}^{3+}$ laser require a concept to reduce parasitic effects due to amplified spontaneous emission and lasing in the Yb$^{3+}$ system. Beside of conventional step index LMA fibers experimets with multi-filament fibers \cite{Kuhn2010} and Er$^{3+}$  doped photonic crystal fibers \cite{Kuhn2011a} were performed and show up new options for high power lasers at $1550\,{\rm nm}$.

At the time of writing it seems likely, that the power levels required for the next generation GWD's laser light sources will be achieved most effectively by the coherent combination of two or more high power lasers. This scheme was tested at different power levels (see ref. \cite{Tuennermann2011} and references therein). Additional highly relevant experience will be gained from the Advanced Virgo and KAGRA GWD high power laser, which rely the coherent summation of several high power systems.

\subsection{Laser stabilization}\label{stabilization}
The electric field of the laser light source of GWDs can be described by
\begin{equation}
{\bm E}({\bm r},t)={\bm e}_p({\bm r},t) \cdot \Re(U({\bm r},t))
\end{equation}
with the scalar complex field amplitude
\begin{equation}
U({\bm r},t)=U_0(t)\cdot \Phi({\bm r},t) \cdot {\rm e}^{{\rm i} (2\uppi \nu t- k z }).
\end{equation}
This equation for a gaussian beam propagating in the $z$-direction includes all laser parameter that have to fulfill special requirement. The polarization state of the light described by ${\bm e}_p$ has to be linear and the spatial beam profile $\Phi({\bm r},t)$ has to be close to the fundamental Gausian mode or a pure higher order Laugerre Gauss mode. Furthermore the laser needs to oscillate in only one longitudinal mode, often called single-frequency operation. Very important is the time evolution of the slowly varying complex field amplitude $U_0(t)$ that defines the laser frequency noise and the laser power noise. Requirements for both quantities are typically given in the frequency domain via limits on the power spectral density. It is convenient to use a normalized  version of the power noise, the relative power noise (RPN). The spectral density description is helpful for the noise propagation through the GWD via transfer functions and for a comparison of the propagated noise with the desired GWD sensitivity. In addition special care has to be taken to keep the non-stationary noise and the glitch rate low. Furthermore the peak-peak variation and drifts of the laser frequency and power have to be low enough to avoid any saturation effects in the sensors or actuators of the active feedback control systems (FCS) used in the power and frequency stabilization.

In past and current GWD generations the spatial variations of the beam shape and the beam pointing fluctuations of the PSL were reduced by passive filtering with rigidly assembled optical cavities called pre-mode-cleaner. These devices furthermore provide power noise filtering at high frequencies
\linebreak
\vspace*{-2mm}

\begin{figure}[H]
\centering
\includegraphics{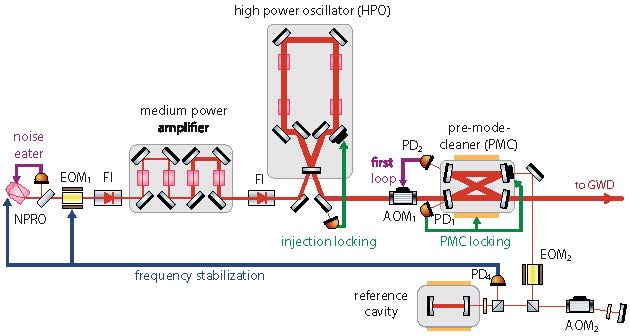} 
\caption{(Color online) Schematic setup of the Advanced LIGO pre-stabilized high power laser. The light of a master laser (NPRO) is phase modulated by an electro-optical modulator (${\rm EOM_1}$) and amplified by a medium power amplifier and an injection locked high power oscillator. Faraday isolators (FI) avoid any disturbance of the NPRO by backwards directed light. An acousto optic modulator (${\rm AOM_1}$) is used to stabilize the laser power and an optical ring cavity called pre-mode-cleaner (PMC) provides spatial filtering and power noise reduction in the MHz frequency range. A fraction of the light leaving the PMC is phase modulated by ${\rm EOM_2}$ and frequency shifted by ${\rm AOM_2}$ and then compared with a resonance frequency of a reference cavity. The reflected light is detected with the photodiode ${\rm PD_3}$ in a Pound-Drever-Hall frequency stabilization scheme with feed-back to the NPRO and the phase-correcting ${\rm EOM_1}$. }
\label{setup_PSL}
\end{figure}

\noindent required for a low noise background at phase modulation frequencies used for the GWD interferometer's length and alignment control. Despite of their positive effect such PMCs cause a conversion of pointing, PMC length noise and laser frequency fluctuations into RPN behind the PMC if not kept perfectly resonant with the injected light field. Hence PMCs have to be kept resonant to the incident laser light within tight boundaries.

Even though the detailed design of next generation GWDs and their laser source is not yet defined it is likely that the laser pre-stabilization will reuse concepts applied to the PSLs of the currently commissioned GWD generation. The PSL of the Advanced LIGO GWD  \cite{Kwee2012} can serve as a concrete example to review these concepts.
Figure \ref{setup_PSL} shows a schematic overview of the Advanced LIGO PSL.

A single frequency, linear polarized laser system with an output power of approximately $150\,{\rm W}$ is used as the laser source. Its frequency is defined by the low noise $2\,{\rm W}$ NPRO  \cite{Innolight} used as the master laser. The NPRO is amplified by a single pass laser amplifier to $35\,{\rm W}$ \cite{Frede2007a}. This medium power amplifier consists of four end-pumped ${\rm{Nd:YVO}}_4$ amplifier crystals.
A second amplification stage uses the injection-locking technique.  A four head Nd:YAG high power ring oscillator is locked to the $35\,{\rm W}$ stage to generate a $150\,{\rm W}$ beam with the center frequency and frequency stability defined by the NPRO.
(For a detailed description of this laser source see ref. \cite{Winkelmann2011}.)

From a stabilization point of view the unstabilized noise (free running noise) of the laser system and the actuators  provided by the laser system for the control of these fluctuations are the most important laser properties. Even with a very low noise laser design it is almost certain, that the free running noise will be too high for a direct injection of the light into next generation GWDs. Hence the laser source should provide fast frequency and power actuators with a large range. External devices like acousto-optic or electro-optic modulator can be used to increase the actuation bandwidth and range. This approach is independent of the laser design but requires that those devices can withstand the high optical power and do not significantly degrade the spatial beam profile.

It is likely that spatial filtering of the laser beam by a PMC is required before the beam is injected into the GWD. This filtering includes a reduction of the HOM content of the laser beam as well as its beam pointing (lateral or angular fluctuations of the beam's propagation direction). Furthermore power noise filtering is provided by a PMC at high frequencies where active power stabilization becomes problematic. From a practical point of view it is advisable to pick-off sample beams for the power and frequency stabilization downstream of the PMC similar to the example give in Figure \ref{setup_PSL}. The advantage of this approach is that the spatial properties of those pick-off beams are independent of potential drifts of the laser. Past experience has shown that it is extremely important to install the PMC in a clean environment to avoid an accumulation of contamination on the PMC mirrors. This could cause thermal effects like a change in the mirror's radius of curvature and a reduction of the transmitted power. Depending on the global frequency control scheme a large actuation range for the PMC's length and hence resonance frequency might be required to allow for large laser frequency changes without PMC lock losses.

A seismically isolated rigid-spacer high finesse cavity installed in a vacuum vessel was used as a frequency reference for GWD PSLs so far. The Pound-Drever-Hall (PDH) method \cite{Drever1983, Black2001} is used to generate an electronic signal that represents the frequency difference $\Delta \nu _{\rm{RC}}$ between the laser and a resonance frequency of this cavity. After appropriate filtering this signal is fed back to the frequency control actuators of the laser. In the Advanced LIGO case these are the thermal and PZT actuator of the NPRO and a fast phase correcting EOM (see ${\rm {EOM}_1}$ in Figure \ref{setup_PSL}). The reference cavity is part of a complicated neested length and frequency control system of the GWD with several frequency references. Hence it is not possible to predict the role of rigid spacer reference cavities in future generation GWDs. It might be possible to design a control scheme with the PMC as the PSL's frequency reference, which would make a dedicated reference cavity redundant.

Since radiation pressure is an important coupling path of laser RPN into the GW readout channel
the PSL RPN requirements for next generation GWDs are largely independent of the laser design.
It is likely, that next generation GWDs require that the power fluctuations be below $\rm{RPN} = 1\!\times\!10^{-9}/\sqrt{\rm{Hz}}$. Such a high stability can only be achieved by the use of a photodiode array similar to the one tested in ref. \cite{Kwee2009a} or by a sensing technique known as optical AC coupling \cite{Kwee2008b, Kwee2009, Kwee2011}. The final power noise sensors have to be placed downstream of the IMC such that the power stabilization FCS can compensate pointing to RPN coupling at the IMC. Currently no stand-alone power noise actuator is available that can tolerate several hundred Watts of laser power and does not degrade the spatial beam profile. Therefore GWD power control loops need to feed back to the PSL's power actuator. The IMC power modulation transfer function (a low pass with a corner frequency equal to the IMC's half linewidth) reduces the achievable unity gain frequency of the final power stabilization FCS such that a large power noise reduction is required inside the PSL subsystem already.

As mentioned above it is advisable to use a beam sample picked off after the PMC for the PSL power stabilization. This has the additional advantage, that the beam pointing on the power stabilization photodiode and hence its coupling via photodiode inhomogeneities the detected photocurrent is reduced. This effect was responsible for one of the most important noise sources in the Advanced LIGO PSL power stabilization. Scattered light was identified as a second important disturbance in the power stabilization such that a well designed  scattered light control is advisable for next generation PSLs.

Even though the laser and stabilization concepts discussed in this paper are valid options for next generation GWDs we would like to point out that future laser research will be strongly influenced by design choices to be made for the optical design of next generarion GWDs. Hence further improvements of existing PSL concepts or totally new developments might become necessary.

\vspace*{2mm} \Acknowledgements{\bahao The authors would like to thank KITPC for its financial support during The Next Detectors for Gravitational Wave Astronomy workshop in Beijing in 2015.}

\end{multicols}

\end{document}